\documentclass{aastex631}

\usepackage{placeins}

\shorttitle{Neutrino Emission Associated with IceCube’s Highest-Energy Tracks}
\shortauthors{IceCube Collaboration et al.}
\graphicspath{{./}{figures/}}

\begin{document}

\title{Search for Continuous and Transient Neutrino Emission Associated with IceCube's Highest-Energy Tracks: An 11-Year Analysis}

\email{analysis@icecube.wisc.edu}

\affiliation{III. Physikalisches Institut, RWTH Aachen University, D-52056 Aachen, Germany}
\affiliation{Department of Physics, University of Adelaide, Adelaide, 5005, Australia}
\affiliation{Dept. of Physics and Astronomy, University of Alaska Anchorage, 3211 Providence Dr., Anchorage, AK 99508, USA}
\affiliation{Dept. of Physics, University of Texas at Arlington, 502 Yates St., Science Hall Rm 108, Box 19059, Arlington, TX 76019, USA}
\affiliation{CTSPS, Clark-Atlanta University, Atlanta, GA 30314, USA}
\affiliation{School of Physics and Center for Relativistic Astrophysics, Georgia Institute of Technology, Atlanta, GA 30332, USA}
\affiliation{Dept. of Physics, Southern University, Baton Rouge, LA 70813, USA}
\affiliation{Dept. of Physics, University of California, Berkeley, CA 94720, USA}
\affiliation{Lawrence Berkeley National Laboratory, Berkeley, CA 94720, USA}
\affiliation{Institut f{\"u}r Physik, Humboldt-Universit{\"a}t zu Berlin, D-12489 Berlin, Germany}
\affiliation{Fakult{\"a}t f{\"u}r Physik {\&} Astronomie, Ruhr-Universit{\"a}t Bochum, D-44780 Bochum, Germany}
\affiliation{Universit{\'e} Libre de Bruxelles, Science Faculty CP230, B-1050 Brussels, Belgium}
\affiliation{Vrije Universiteit Brussel (VUB), Dienst ELEM, B-1050 Brussels, Belgium}
\affiliation{Department of Physics and Laboratory for Particle Physics and Cosmology, Harvard University, Cambridge, MA 02138, USA}
\affiliation{Dept. of Physics, Massachusetts Institute of Technology, Cambridge, MA 02139, USA}
\affiliation{Dept. of Physics and The International Center for Hadron Astrophysics, Chiba University, Chiba 263-8522, Japan}
\affiliation{Department of Physics, Loyola University Chicago, Chicago, IL 60660, USA}
\affiliation{Dept. of Physics and Astronomy, University of Canterbury, Private Bag 4800, Christchurch, New Zealand}
\affiliation{Dept. of Physics, University of Maryland, College Park, MD 20742, USA}
\affiliation{Dept. of Astronomy, Ohio State University, Columbus, OH 43210, USA}
\affiliation{Dept. of Physics and Center for Cosmology and Astro-Particle Physics, Ohio State University, Columbus, OH 43210, USA}
\affiliation{Niels Bohr Institute, University of Copenhagen, DK-2100 Copenhagen, Denmark}
\affiliation{Dept. of Physics, TU Dortmund University, D-44221 Dortmund, Germany}
\affiliation{Dept. of Physics and Astronomy, Michigan State University, East Lansing, MI 48824, USA}
\affiliation{Dept. of Physics, University of Alberta, Edmonton, Alberta, Canada T6G 2E1}
\affiliation{Erlangen Centre for Astroparticle Physics, Friedrich-Alexander-Universit{\"a}t Erlangen-N{\"u}rnberg, D-91058 Erlangen, Germany}
\affiliation{Physik-department, Technische Universit{\"a}t M{\"u}nchen, D-85748 Garching, Germany}
\affiliation{D{\'e}partement de physique nucl{\'e}aire et corpusculaire, Universit{\'e} de Gen{\`e}ve, CH-1211 Gen{\`e}ve, Switzerland}
\affiliation{Dept. of Physics and Astronomy, University of Gent, B-9000 Gent, Belgium}
\affiliation{Dept. of Physics and Astronomy, University of California, Irvine, CA 92697, USA}
\affiliation{Karlsruhe Institute of Technology, Institute for Astroparticle Physics, D-76021 Karlsruhe, Germany }
\affiliation{Karlsruhe Institute of Technology, Institute of Experimental Particle Physics, D-76021 Karlsruhe, Germany }
\affiliation{Dept. of Physics, Engineering Physics, and Astronomy, Queen's University, Kingston, ON K7L 3N6, Canada}
\affiliation{Department of Physics {\&} Astronomy, University of Nevada, Las Vegas, NV, 89154, USA}
\affiliation{Nevada Center for Astrophysics, University of Nevada, Las Vegas, NV 89154, USA}
\affiliation{Dept. of Physics and Astronomy, University of Kansas, Lawrence, KS 66045, USA}
\affiliation{Centre for Cosmology, Particle Physics and Phenomenology - CP3, Universit{\'e} catholique de Louvain, Louvain-la-Neuve, Belgium}
\affiliation{Department of Physics, Mercer University, Macon, GA 31207-0001, USA}
\affiliation{Dept. of Astronomy, University of Wisconsin{\textendash}Madison, Madison, WI 53706, USA}
\affiliation{Dept. of Physics and Wisconsin IceCube Particle Astrophysics Center, University of Wisconsin{\textendash}Madison, Madison, WI 53706, USA}
\affiliation{Institute of Physics, University of Mainz, Staudinger Weg 7, D-55099 Mainz, Germany}
\affiliation{Department of Physics, Marquette University, Milwaukee, WI, 53201, USA}
\affiliation{Institut f{\"u}r Kernphysik, Westf{\"a}lische Wilhelms-Universit{\"a}t M{\"u}nster, D-48149 M{\"u}nster, Germany}
\affiliation{Bartol Research Institute and Dept. of Physics and Astronomy, University of Delaware, Newark, DE 19716, USA}
\affiliation{Dept. of Physics, Yale University, New Haven, CT 06520, USA}
\affiliation{Columbia Astrophysics and Nevis Laboratories, Columbia University, New York, NY 10027, USA}
\affiliation{Dept. of Physics, University of Oxford, Parks Road, Oxford OX1 3PU, United Kingdom}
\affiliation{Dipartimento di Fisica e Astronomia Galileo Galilei, Universit{\`a} Degli Studi di Padova, 35122 Padova PD, Italy}
\affiliation{Dept. of Physics, Drexel University, 3141 Chestnut Street, Philadelphia, PA 19104, USA}
\affiliation{Physics Department, South Dakota School of Mines and Technology, Rapid City, SD 57701, USA}
\affiliation{Dept. of Physics, University of Wisconsin, River Falls, WI 54022, USA}
\affiliation{Dept. of Physics and Astronomy, University of Rochester, Rochester, NY 14627, USA}
\affiliation{Department of Physics and Astronomy, University of Utah, Salt Lake City, UT 84112, USA}
\affiliation{Oskar Klein Centre and Dept. of Physics, Stockholm University, SE-10691 Stockholm, Sweden}
\affiliation{Dept. of Physics and Astronomy, Stony Brook University, Stony Brook, NY 11794-3800, USA}
\affiliation{Dept. of Physics, Sungkyunkwan University, Suwon 16419, Korea}
\affiliation{Institute of Physics, Academia Sinica, Taipei, 11529, Taiwan}
\affiliation{Dept. of Physics and Astronomy, University of Alabama, Tuscaloosa, AL 35487, USA}
\affiliation{Dept. of Astronomy and Astrophysics, Pennsylvania State University, University Park, PA 16802, USA}
\affiliation{Dept. of Physics, Pennsylvania State University, University Park, PA 16802, USA}
\affiliation{Dept. of Physics and Astronomy, Uppsala University, Box 516, S-75120 Uppsala, Sweden}
\affiliation{Dept. of Physics, University of Wuppertal, D-42119 Wuppertal, Germany}
\affiliation{Deutsches Elektronen-Synchrotron DESY, Platanenallee 6, 15738 Zeuthen, Germany }

\author[0000-0001-6141-4205]{R. Abbasi}
\affiliation{Department of Physics, Loyola University Chicago, Chicago, IL 60660, USA}

\author[0000-0001-8952-588X]{M. Ackermann}
\affiliation{Deutsches Elektronen-Synchrotron DESY, Platanenallee 6, 15738 Zeuthen, Germany }

\author{J. Adams}
\affiliation{Dept. of Physics and Astronomy, University of Canterbury, Private Bag 4800, Christchurch, New Zealand}

\author[0000-0002-9714-8866]{S. K. Agarwalla}
\altaffiliation{also at Institute of Physics, Sachivalaya Marg, Sainik School Post, Bhubaneswar 751005, India}
\affiliation{Dept. of Physics and Wisconsin IceCube Particle Astrophysics Center, University of Wisconsin{\textendash}Madison, Madison, WI 53706, USA}

\author[0000-0003-2252-9514]{J. A. Aguilar}
\affiliation{Universit{\'e} Libre de Bruxelles, Science Faculty CP230, B-1050 Brussels, Belgium}

\author[0000-0003-0709-5631]{M. Ahlers}
\affiliation{Niels Bohr Institute, University of Copenhagen, DK-2100 Copenhagen, Denmark}

\author[0000-0002-9534-9189]{J.M. Alameddine}
\affiliation{Dept. of Physics, TU Dortmund University, D-44221 Dortmund, Germany}

\author{N. M. Amin}
\affiliation{Bartol Research Institute and Dept. of Physics and Astronomy, University of Delaware, Newark, DE 19716, USA}

\author[0000-0001-9394-0007]{K. Andeen}
\affiliation{Department of Physics, Marquette University, Milwaukee, WI, 53201, USA}

\author[0000-0003-2039-4724]{G. Anton}
\affiliation{Erlangen Centre for Astroparticle Physics, Friedrich-Alexander-Universit{\"a}t Erlangen-N{\"u}rnberg, D-91058 Erlangen, Germany}

% ...

\date{\today}

\collaboration{406}{IceCube Collaboration}



\begin{abstract}

IceCube alert events are neutrinos with a moderate-to-high probability of having astrophysical origin. In this study, we analyze 11 years of IceCube data and investigate 122 alert events and a selection of high-energy tracks detected between 2009 and the end of 2021. This high-energy event selection (alert events + high-energy tracks) has an average probability of $\geq 0.5$ to be of astrophysical origin. We search for additional continuous and transient neutrino emission within the high-energy events' error regions. We find no evidence for significant continuous neutrino emission from any of the alert event directions. The only locally significant neutrino emission is the transient emission associated with the blazar TXS~0506+056, with a local significance of $ 3 \sigma$, which confirms previous IceCube studies. When correcting for 122 test positions, the global p-value is $0.156$ and is compatible with the background hypothesis. We constrain the total continuous flux emitted from all 122 test positions at 100~TeV to be below $1.2 \times 10^{-15}$~(TeV cm$^2$ s)$^{-1}$ at 90\% confidence assuming an $E^{-2}$ spectrum. This corresponds to 4.5\% of IceCube's astrophysical diffuse flux. Overall, we find no indication that alert events, in general, are linked to lower-energetic continuous or transient neutrino emission.

\end{abstract}

\keywords{Neutrino Astronomy (1100) --- High-energy astrophysics (739) --- Transient sources (1851) --- Blazars (164) --- Active galactic nuclei (16) \vspace{1cm}}


\section{Introduction} \label{sec:intro}

The IceCube Neutrino Observatory \citep{icecube_collaboration_2017} is a Cherenkov detector using a cubic kilometer of Antarctic ice at the geographic South Pole to primarily (but not exclusively) study high-energy astrophysical neutrinos. Its duty cycle is greater than 99\% \citep{icecube_collaboration_2017}, and its field of view covers the full sky while being most sensitive to high-energy neutrino events near the celestial equator. This makes IceCube ideal for surveying the sky \citep{2017APh....92...30A}. As part of the realtime program, IceCube alerts other telescopes upon detection of a neutrino event with a high probability of being of astrophysical origin, which can then trigger follow-up observations \citep{2017APh....92...30A, Blaufuss:20199c, Kintscher_2016}. 

On the 22nd of September 2017, IceCube detected a neutrino of likely astrophysical origin (IceCube-170922A\footnote{\url{https://gcn.gsfc.nasa.gov/notices_amon/50579430_130033.amon}}). This triggered multi-wavelength follow-up observations, which detected a flaring blazar (TXS~0506+056) at the reconstructed origin direction of IceCube-170922A \citep{2018Sci...361.1378I}. This correlation is significant at a $3 \sigma$ level \citep{2018Sci...361.1378I}. Additionally, a neutrino flare was identified originating from the same direction between September 2014 and March 2015 with a significance of $3.5 \sigma$ \citep{2018Sci...361..147I}. 

This detection demonstrates that IceCube alerts can point to neutrino source candidates due to their high probability of being of astrophysical origin, and we aim to investigate the origin directions of other IceCube alerts. A preliminary search showed no indication of continuous neutrino emission \citep{2019ICRC...36..929K}. However, the IceCube alert criteria have since been updated \citep{Blaufuss:20199c, abbasi2023icecat1}. The IceCube data have also been reprocessed with improved calibration of the optical sensors \citep{PhysRevD.104.022002, Aartsen_2020}. This leads to improved energy and direction reconstruction compared to previous results in \citet{2018Sci...361..147I, 2021arXiv210109836I, 2020time_integrated_neutrino_sources, aartsen2015searches, 2021ApJ...911...67A, 2021ApJ...920L..45A}. A first analysis benefiting from this new processing \citep{doi:10.1126/science.abg3395} detected the neutrino signal from the Seyfert II galaxy NGC~1068 with a significance of 4.2$\sigma$ (compared to 2.9$\sigma$ in \citet{2020time_integrated_neutrino_sources}). A large part of the increase ($0.9\sigma$) is due to improved data processing and calibration. More details about effects on data are discussed in appendix \ref{appendix:transient_source_analysis} and in the supplementary material of \citet{doi:10.1126/science.abg3395}.

In this work, we analyze 11 years of reprocessed IceCube data (through-going muon tracks, see Table \ref{tab:data_sample}) and search for an excess of neutrino-induced muons. We apply a conservative lower limit on the angular uncertainty of 0.2 degrees, whereas the median angular resolution is 0.57 degrees (compared to a median angular resolution of 0.59 degrees before the reprocessing). We identify possible neutrino production sites by looking at the origin of high-energy neutrinos that have a high probability of being of astrophysical origin. IceCube's highest energy neutrinos with the largest astrophysical purity are events from the new selection of IceCube alerts published in the so-called ``gold'' alert channel \citep{Blaufuss:20199c, abbasi2023icecat1}. Additionally, we extend the list by including 18 high-energy events from \citet{2022ApJ...928...50A} that were confirmed to be likely astrophysical events by a new event classifier \citep{kronmueller2019application}. Since we use a combination of IceCube alert events and high-energy tracks identified retrospectively, we will refer to our event selection as ``alert+ events'' for brevity. All IceCube data used in this work (lower and high-energy events) have been reprocessed.

\begin{figure}
    \centering
    \includegraphics[width=0.7\textwidth]{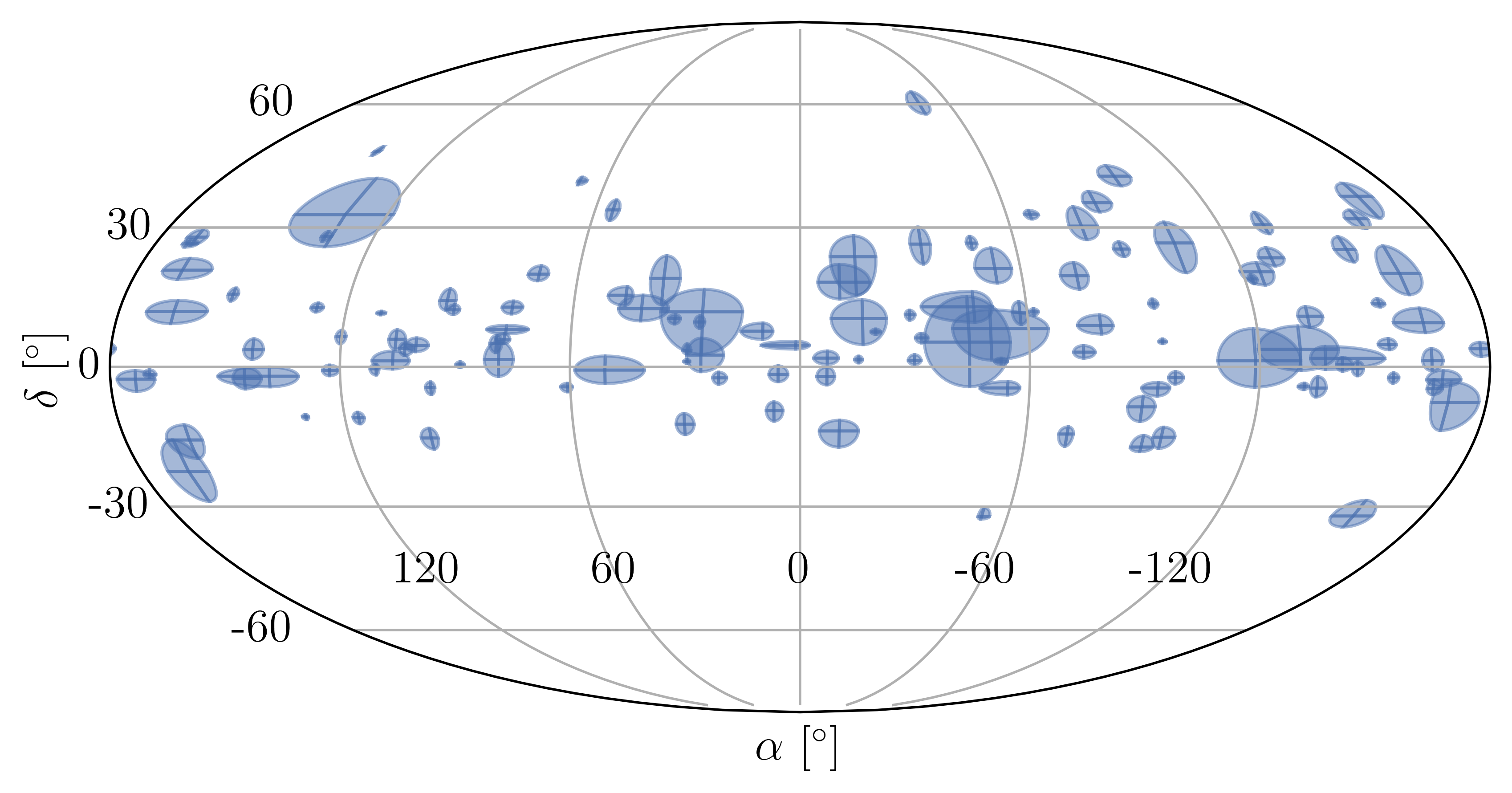}
    \caption{Sky map in right ascension and declination (epoch=J2000) with the arrival directions of events fulfilling the IceCube alert criteria (with the highest probability to be of astrophysical origin) we investigate in this work. The events were detected between August 2009 and the end of 2021. The shaded regions represent the 90\% uncertainty region of the reconstruction.}
    \label{fig:skymap}
\end{figure}

In this work, we excluded alert+ events within $30$~degrees of the geographic poles (affecting three events), for which we have smaller statistics for the background. Other IceCube analyses have applied different declination cuts (for example including all events up to $81^{\circ}$ declination \citep{doi:10.1126/science.abg3395} or up to $82^{\circ}$ \citep{2020time_integrated_neutrino_sources}). Additionally, we removed alert+ events with large uncertainties ($\geq 100 ~\text{square degrees}$, affecting two events). As a result, our final sample consists of 122 high-energy events (104 IceCube alert events and 18 high-energy tracks, listed in Table \ref{tab:alert table}), detected between 2009 and the end of 2021. On average, our selected alert+ events have a probability to be astrophysical of $\gtrsim 0.5$. The probability to be astrophysical is spectrum dependent and based on the muon neutrino spectrum measured by IceCube \citep{Haack:2017E1, 2022ApJ...928...50A}. The median angular resolution (90\% uncertainty regions) of alert+ events is 2.1 degrees. 
In Figure \ref{fig:skymap}, we show a map of all arrival directions and their 90\% uncertainty regions of IceCube alert+ events investigated in this work. These events provide positions of interest analogous to a catalog of possible neutrino sources. Since IceCube alert+ events trigger this analysis, we remove the respective alert+ event from the 11 years of IceCube data when running the analysis. We present the analysis method in Section \ref{sec:method} and the results in Section \ref{sec:results}.

\begin{table}[h]
    \centering
    \caption{Overview of the improved and reprocessed data samples in this analysis. The columns list the configuration of the detector (``IC''  and the number of deployed strings), the uptime (livetime) of the detector in days, the number of events in each sample, and the start and end dates of the data subset. }
    \begin{tabular}{ccccc}
    \tableline
    Year &Livetime [days] &Number of events &Start & End \\
     \tableline
    IC59 &353.578 &107011 & 2009 May 5 & 2010 May 31\\
    IC79 &316.045 &93133 &2010 June 1 & 2011 May 13 \\
    IC86 2011-2019 &3184.163 &1133364 & 2011 May 13 & 2020 May 29 \\
    \tableline
    \end{tabular}
    \label{tab:data_sample}
\end{table}

\section{Analysis Method}\label{sec:method}

We use an unbinned likelihood approach as presented by \citet{Braun:2008bg}. In this work, we investigate two source types: continuous sources and transient sources. %
We compare two hypotheses (each with a set of parameters $\vec{\theta}$)
\begin{itemize}
    \item \textbf{Background hypothesis} $H_0(\vec{\theta}_0)$: The background comprises atmospheric neutrinos, atmospheric muons (remaining after event selection cuts), and diffuse astrophysical neutrinos. The flux is uniform in time and right ascension.\\
    
    \item \textbf{Signal hypothesis} $H_1(\vec{\theta}_1)$: There is a signal component additional to the atmospheric background and the average diffuse astrophysical neutrino emission. The signal neutrinos cluster around their source (subscript $S$) at right ascension, declination $\vec{x}_S = (\alpha_S, \delta_S)$. The energy spectrum of the emitted flux is an unbroken power law: $\frac{d \phi}{d E_{\nu}} \propto E_{\nu}^{-\gamma}$. In the specific case of a transient source hypothesis (see Section \ref{sec:analysis:transientSources}), the neutrino emission has a Gaussian time profile with mean $\mu_T$ and width $\sigma_T$.\\
    
\end{itemize}

We remove the high-energy alert+ events that triggered this analysis from the data set. Hence, we look for additional neutrino emission from the direction of the high-energy alert+ events. 
We then maximize the likelihood, $\mathcal{L}$, and compute the likelihood ratio
\begin{equation}\label{eq:ts_general_1}
    \lambda(\vec{x}) = \frac{\sup \mathcal{L}(H_0)}{\sup \mathcal{L}(H_1)}.
\end{equation}
The likelihood maximization varies the expectation value of the number of detected signal neutrinos, $n_S$, and the emitted energy spectral index, $\gamma$. We allow values for $\gamma$ between 1.5 and 4. For the background hypothesis, $n_S$ is fixed to 0. 

The likelihood is the probability density of observing the data given a specific hypothesis. The probability density of observing an event, $i$, is a sum of its probability to be signal, $S_i$, or background, $B_i$: $\frac{n_S}{N} S_i + \left(1 - \frac{n_S}{N} \right) B_i$, where $N$ is the total number of detected events (signal and background combined). 

We define the test statistic, $\text{TS}$, as 
\begin{equation}\label{eq:TS_timeint_exp}
\text{TS} = -2 \ln \lambda = -2\ln  \left[ \frac{\mathcal{L} \left( \hat{\vec{\theta}}_0|\vec{x}_S \right) }{\mathcal{L} \left( \hat{\vec{\theta}}_1|\vec{x}_S \right) } \right] = 2 \ln  \left[\frac{\mathcal{L}\left( n_S=\hat{n}_S \right) }{\mathcal{L} \left( n_S=0 \right) } \right] 
 = 2 ~ \sum_{i} \ln \left[\frac{\hat{n}_S}{N}\left(\frac{S_i}{B_i}-1\right) +1\right],
\end{equation}
for a signal hypothesis with the best-fit value of $\hat{n}_S$ neutrinos (the ``\,$\hat{}$\,'' denoting the best-fit of a parameter) as the mean number of neutrinos we expect to detect from the neutrino source.

The investigated source candidates have directional uncertainties (see Figure \ref{fig:skymap}). However, we assume potential sources are smaller than the best resolution of 0.2 degrees in our data (TXS~0506+056 has an angular size of $\sim 2.6 $ arc-seconds). Hence, we fit the best point-source position within a reconstructed 90\% uncertainty region by dividing the region in a grid with steps of $0.2^{\circ}$ in right ascension and declination, the best angular uncertainty for events used in this study. The likelihood is optimized at each grid point. The grid point yielding the best result (i.e., the highest TS value) is subsequently considered the point-source position. 

This procedure is run on different realizations of background data $\sim 10^4$ times. The background data are all 11 years of muon tracks with randomly assigned right ascensions. In the final step, we calculate the test statistic value, $\text{TS}_{\text{data}}$, for the true data and compare this with the simulated background test statistic distribution. The local p-value is the probability of getting this $\text{TS}_{\text{data}}$ (or a larger value) from a random background realization. This procedure is repeated for all remaining regions in the sky, yielding 122 local p-values. From these 122 values, we take the most significant local p-value, $p_{0}$, to identify the most significant source.

As a next step, we correct the significance for having tested 122 regions in the sky. Considering only background realizations, we take the most significant p-value out of 122 positions for each realization and generate a distribution of best local p-values, $p_{0, \text{BG}}$. The final global p-value of our analysis is the probability for $p_{0, \text{BG}}$ to be at least as significant as the p-value we got from our real data, $p_0$. Since we are investigating only a limited number of points (122), weaker neutrino emissions have a higher significance in this analysis than in an all-sky scan, for example, in \citet{doi:10.1126/science.abg3395}.

When testing the method with Monte Carlo simulations, the best-fit number of signal neutrinos, $\hat{n}_S$, and source spectral index, $\hat{\gamma}$, show a bias compared to the true simulated source properties. For sources with simulated hard spectral indices (i.e., $\gamma = 2$), there is a tendency to fit slightly softer spectra and a slightly larger number of signal neutrinos. For simulated sources following softer spectral indices (i.e., $\gamma = 3$), the tendency is reversed to fitting slightly harder spectral indices and smaller numbers of signal neutrinos. Appendix \ref{appendix:bias} presents a more in-depth discussion of this bias. Correcting the bias is not straightforward, and we have decided not to include an, at best, incomplete correction. Hence, the best-fit fluxes are only indicative. This bias does not affect the flux limits since they are based on simulated fluxes where the true source strength is known. 

\subsection{Time-integrated search for continuous sources}
We define the signal and background probability density functions (pdfs) $S_i$ and $B_i$ in a spatial and an energy part (see, e.g., \citet{Braun:2008bg, Abbasi_2011}). The spatial part depends on the source position $\vec{x}_S$ and the reconstructed event properties: reconstructed origin $\vec{x}_i$ and the angular uncertainty of the reconstructed origin $\sigma_i$. The energy part depends on the reconstructed muon energy, $E_i$, the reconstructed origin declination, $\delta_i$, and the source energy spectral index, $\gamma$. The signal pdf for a steady source is hence 
\begin{equation}\label{eq:signal_timeint_gen}
S_i (\vec{x}_i ,  E_i | \sigma_i, \vec{x}_S, \gamma) = S_{\text{spatial}} (\vec{x}_i| \sigma_i , \vec{x}_S) \cdot S_{\text{energy}} (E_i | \delta _i , \gamma ) 
= \frac{1}{2 \pi \sigma_i ^2} \exp \left( \frac{- | \vec{x}_i - \vec{x}_S | ^2}{2 \sigma _i ^2} \right) \cdot S_{\text{energy}} (E_i | \delta _i, \gamma ) .
\end{equation}
The energy pdf, $S_{\text{energy}}$, is the probability of detecting a neutrino with reconstructed energy, $E_i$, at declination, $\delta_i$, assuming the source emits neutrinos with a spectrum of $E^{-\gamma}$. The background pdfs, $B_i$, are defined similarly 
\begin{equation}\label{eq:bg_timeint_exp}
B_i \left( \vec{x}_i, E_i \right) = B_{\text{spatial}} (\vec{x}_i) \cdot B_{\text{energy}} (E_i| \delta _i) = \frac{1}{2 \pi} \cdot P(\delta _i) \cdot B_{\text{energy}}(E_i| \delta _i).
\end{equation}
The spatial term depends only on the event declination, $\delta_i$. We assume uniformity in right ascension for the background data due to IceCube's unique position at the South Pole. $B_{\text{energy}}$ is derived directly from experimental data.

Searching for neutrino counterparts of the alert+ events, we want to be sensitive to a single strong emission from one (or a few) sources and, additionally, to faint emissions from a larger number of sources. Hence, our search for continuous sources consists of two parts. The first part searches for single strong neutrino emitters. The second part investigates the overall neutrino emission from all 122 positions of interest. In the latter case, we combine the neutrino emission and define a new test statistic value, $\text{TS}_{\text{stacked}}$, by summing the test statistic values of all alert+ positions, $k$,
\begin{equation}\label{eq:ts_stacking}
    \text{TS}_{\text{stacked}} = \sum _{k} \text{TS}_{k} .
\end{equation}

We take the TS$_k$ from the individual search, hence we do not correct for overlapping uncertainty regions of alert+ events. 

\subsection{Transient sources}\label{sec:analysis:transientSources}
For transient sources, we multiply a temporal pdf with the previously defined spatial and energy pdfs in equations (\ref{eq:signal_timeint_gen}) and (\ref{eq:bg_timeint_exp}) \citep{2010APh....33..175B}. We assume a Gaussian-shaped time profile centered around $\mu_T$ with width $\sigma_T$ for the signal part. The temporal signal pdf becomes

\begin{equation}\label{eq:sig_pdf_temporal_gauss}
S_{T}(t_i | \mu_T, \sigma_T) = \frac{1}{{\sigma_T \sqrt {2\pi } }}\exp \left( - \frac{ \left( {t_i - \mu_T } \right)^2 } {2\sigma_T ^2 } \right),
\end{equation}
with $t_i$ as the time the event was detected. The background expectation is a constant rate over the whole data taking time, $t_{\text{data}}$:

\begin{equation}\label{eq:bkg_pdf_specific}
B_{T} = \frac{1}{\rm{t_{\text{data}}}}.
\end{equation}

The search for time-dependent sources adds another optimization step for the best flaring time. This introduces a bias towards shorter flares since the number of possible shorter flares is larger than the number of possible longer flares. We correct for this effect by multiplying the test statistic by a marginalization factor, $\frac{\sqrt{2 \pi} \sigma_T}{300~\rm{days}}$ \citep{2010APh....33..175B}. Here, 300~days is the maximal flaring time. Longer time scales would result in worse sensitivity than the time-integrated search. We assume a minimal $\sigma_T$ of 5~days to ensure the background uniformity in right ascension. 

Conventional methods to find neutrino flares as in \citet{aartsen2015searches, 2021ApJ...911...67A, 2018Sci...361..147I, 2021ApJ...920L..45A} apply a brute-force scan of all possible time-intervals between events where the ratio of Equation (\ref{eq:signal_timeint_gen}) over Equation (\ref{eq:bg_timeint_exp}) exceeds a certain threshold. This is computationally expensive. The computational cost can be reduced by increasing the required threshold and hence reducing the possible number of intervals scanned. We want to include as few biases as possible, and if following conventional approaches, we would apply the same threshold as in \citet{2018Sci...361..147I}, where the ratio had to be $\geq 1$. However, \citet{2018Sci...361..147I} performs this search only on one position in the sky. In our case, this would mean scanning the uncertainty region of 122 alert+ events in steps of 0.2 degrees and, at each step, evaluating every possible time window between 5 and 300 days in 11 years for neutrino emission. This proved to be computationally unfeasible. To overcome this problem, we investigated new approaches \citep{karl2021search, mythesis, Karl:2023/b}, which do not rely on thresholds, such as a different test statistic to evaluate if an emission is time-dependent \citep{https://doi.org/10.48550/arxiv.2111.02252} or finding an analytical description of the test statistic such that we would not need to simulate a large number of background and signal models. 

Here, we have applied an unsupervised learning algorithm looking for clustering in data: expectation maximization \citep{10.2307/2984875}. This is the first time we apply expectation maximization on IceCube data and use it to fit the best time of transient neutrino emission.

The procedure is as follows \citep{karl2023fitting}. For a source position to be tested (grid point), we assume a two-component mixture model for the temporal distribution of our data (a neutrino flare in the form of a Gaussian signal and uniform background). As a starting flare, we choose a single very broad flare, extending beyond the whole data-taking period. For each event, we compute the probability of it belonging to the neutrino flare (the membership probability). These probabilities are then used to improve the flare parameters iteratively. In the calculation of the membership probability for event $i$, we include the pdf values for the spatial and energy signal and background pdfs (as in Equation (\ref{eq:signal_timeint_gen}) and Equation (\ref{eq:bg_timeint_exp})) as event weights. The membership probability is: 

\begin{equation}
  P_{i, \text{flare}} = \frac{\frac{n_{\text{flare}}}{N} S_i S_{T}(t_i| \mu_T, \sigma_T)  }
  {\frac{n_{\text{flare}}}{N} S_i S_{T} (t_i | \mu_T, \sigma_T) + (1 - \frac{n_{\text{flare}}}{N}) B_i B_T}
    = \frac{ n_{\text{flare}} \frac{S_i}{B_i} S_{T}(t_i | \mu_T, \sigma_T) } { n_{\text{flare}} \frac{S_i}{B_i} S_{T}(t_i | \mu_T, \sigma_T) + \frac{N - n_{\text{flare}}}{t_{\text{data}}}}, 
        \label{eq:expectation_maximization_expression}
\end{equation}

and at each iteration the mean time, $\mu_T$, and the width, $\sigma_T$, are recalculated using

\begin{equation}
    \mu_T = \frac{\sum_i P_{i, \text{flare}} t_i}{\sum_i P_{i, \text{flare}}},
\end{equation}
and 

\begin{equation}
    \sigma_T = \frac{\sum_i P_{i, \text{flare}} \left( t_i - \mu_T \right)^2 }{\sum_i P_{i, \text{flare}}}.
\end{equation}

The quantity $n_{\text{flare}}$ scales the Gaussian temporal pdf according to the expected number of signal events. However, $n_{\text{flare}}$ is only used to determine $\mu_T$ and $\sigma_T$; $n_S$ is fitted independently once we determine the time pdf of the neutrino flare. We stop the iterations when there is no change in the likelihood in the past 20 iterations or once 500 iterations have been performed.

The signal weight, $S_i/B_i$, depends on the assumed source spectral index, $\gamma$. We want to avoid favoring a specific index; hence we run expectation maximization for different fixed spectral indices, $\gamma_{\text{EM}}$, between 1.5 and 4 in steps of 0.2 \citep{karl2023fitting}. We get an optimized time pdf for each $\gamma_{\text{EM}}$. We then optimize the test statistic as in Equation (\ref{eq:TS_timeint_exp}) with the signal and background pdfs, including the temporal pdfs for each $\gamma_{\text{EM}}$. In this step, we fit $n_S$ and $\gamma$ while keeping the temporal pdf with $\hat{\mu}_T(\gamma_{\text{EM}})$ and $\hat{\sigma}_T(\gamma_{\text{EM}})$ fixed. The flare yielding the highest TS value is then the best-fit flare for this grid point. For each alert+, we repeat this procedure at every grid point in the uncertainty region. The point with the most significant result is then the preferred source location. For the background TS distribution, we shuffled the event times and calculated the new right ascension values based on the event azimuths and the shuffled times. 

\section{Results}\label{sec:results}

\subsection{Continuous sources}\label{sec:results_continuous_sources}
The search for the strongest single continuous source yields a global p-value of 0.98 and is compatible with the background hypothesis. We determine an upper flux limit by simulating neutrino emission with an $E^{-2}$ spectrum. The upper flux limit is the flux for which 90\% of the corresponding test statistic distribution lies above the test statistic value of the strongest single continuous source. We get an upper flux limit (for muon neutrinos and antineutrinos) at 90\% confidence level for the most significant position of $\Phi_{90\%, 100\rm{TeV}}^{\nu_{\mu} + \bar{\nu}_{\mu}, \rm{single}} = 6.9 \times 10^{-17}~(\rm{TeV}~\rm{cm}^2~\rm{s})^{-1}$. In general, the energy-dependent flux, $\Phi(E)$, of this flux limit is $\Phi(E) = \Phi_{90\%, 100\rm{TeV}}^{\nu_{\mu} + \bar{\nu}_{\mu}, \rm{single}} \times \left( \frac{E}{100~\rm{TeV}} \right) ^{-2}$. The acceptance for the simulated flux has a limited range in energy. We define the energy range for the flux limit as the central 90\% quantile of detected simulated events. In this case, we limit the flux from 0.9~TeV to 483~TeV. Table \ref{tab:results_timeint} lists the results for all 122 investigated regions. 

For the combined emission of all sources, we get a p-value of 8\%, which is also compatible with the background hypothesis. We determine the 90\% confidence level upper flux limit, $\Phi_{90\%, 100\rm{TeV}}^{\nu_{\mu} + \bar{\nu}_{\mu}, \rm{stacked}}$, by simulating an increasing number of sources emitting a weak flux, $\phi_1$, corresponding to one neutrino coming from a source at the celestial equator --- IceCube's most sensitive region for detecting neutrinos at the highest energies --- in 11 years ($\phi_1 = 4.502 \times 10^{-18}~(\rm{TeV}~\rm{cm}^2~\rm{s})^{-1}$). We repeat the simulation $\sim 10^4$ times for each combined flux and create a $\text{TS}_{\text{stacked}}$ distribution. Based on this distribution, we determine the combined flux strong enough to yield a higher test statistic value than our result with 90\% probability. 

\begin{figure}
\begin{minipage}{0.5\textwidth}
\includegraphics[width=\textwidth]{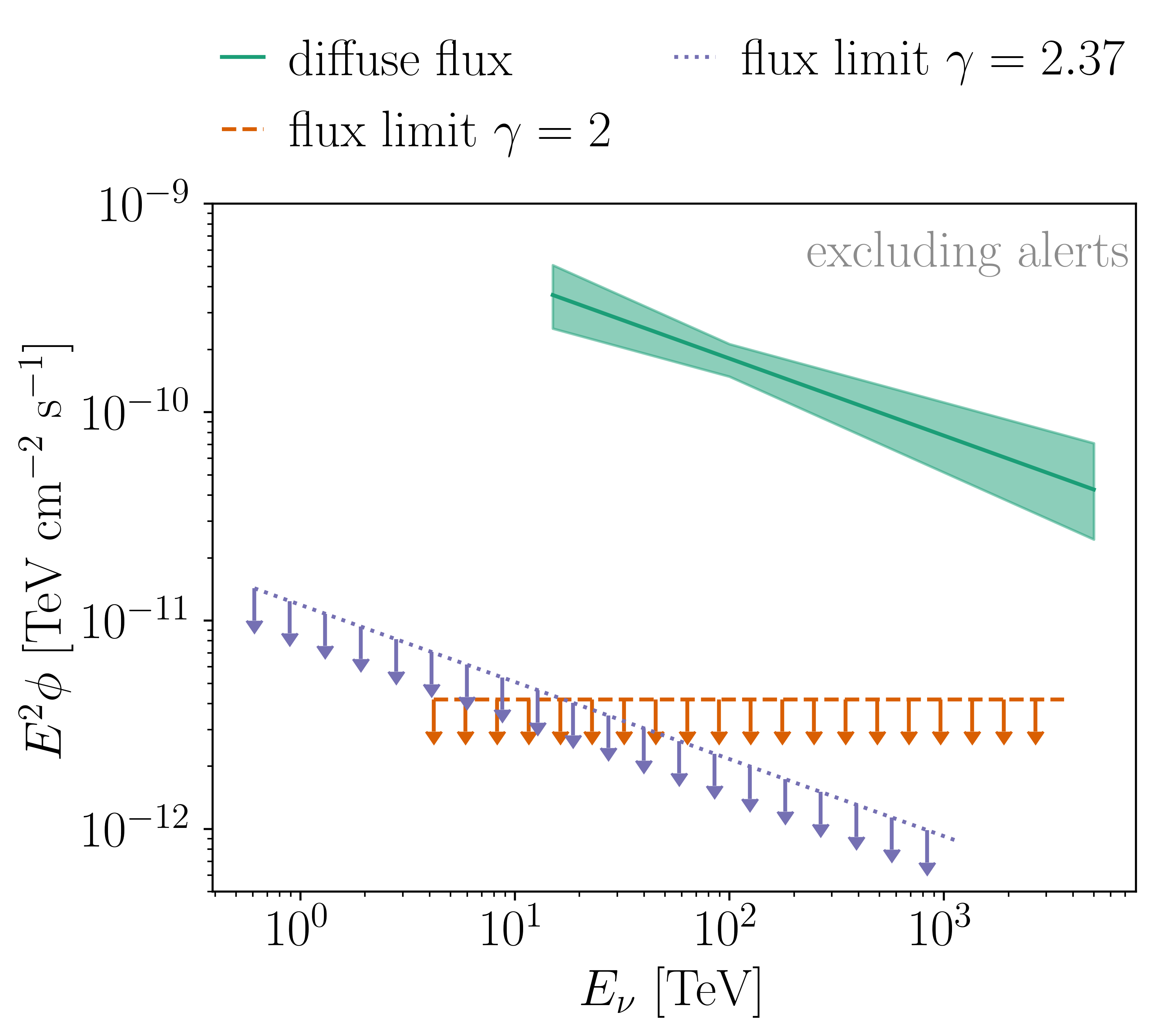}
\end{minipage}
\begin{minipage}{0.5\textwidth}
\includegraphics[width=\textwidth]{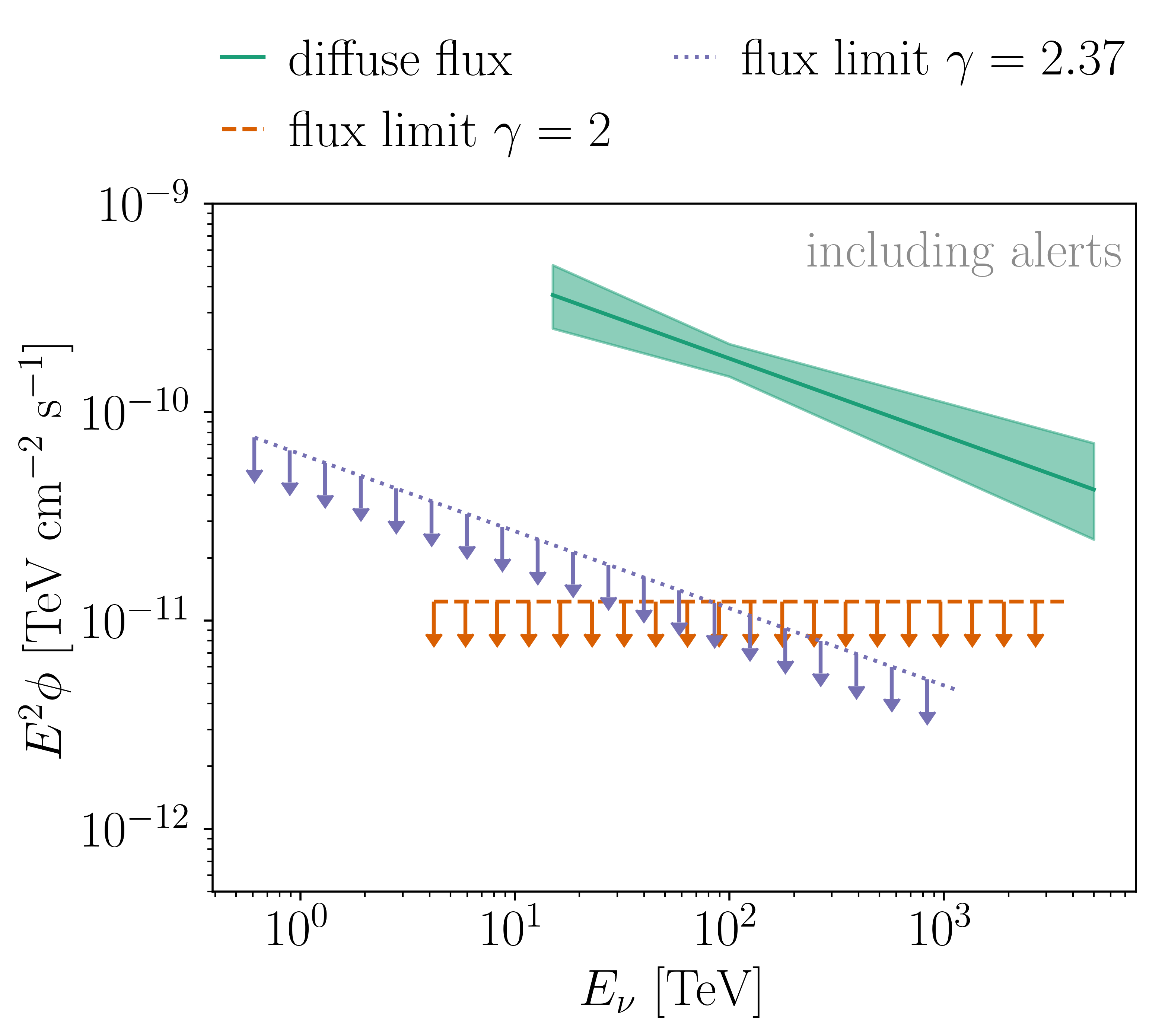}
\end{minipage}
\caption{90\% confidence level upper flux limits assuming for all source candidates combined (dashed orange line) valid in the energy range of 4.2~TeV to 3.6~PeV and a neutrino emission following $E^{-2}$. The green line is the diffuse astrophysical neutrino flux ($\Phi_{\text{diffuse},100\rm{TeV}} = 1.44 \times 10^{-15}  \cdot 4 \pi~(\rm{TeV}~\rm{cm}^2~\rm{s})^{-1})$ in the range of 15~TeV to 5~PeV \citep{2022ApJ...928...50A}). The dotted purple line shows the 90\% confidence level upper flux limit for the spectral index of the diffuse flux ($\gamma = 2.37$) between 0.6~TeV and 1~PeV. \textbf{Left:} The upper flux limit, excluding the alert+ events in the analyzed data, is 1.6\% ($\gamma = 2$) of the astrophysical diffuse flux in the overlapping energy range, and 1.5\% when assuming the same spectral index ($\gamma = 2.37$) as for the astrophysical diffuse flux. \textbf{Right:} The upper flux limit, including the alert+ events in the data, is 4.5\% of the astrophysical diffuse flux in the overlapping energy range for $\gamma = 2$, and 8\% of the diffuse flux when assuming the same spectral index ($\gamma = 2.37$) as for the astrophysical diffuse flux.}
\label{fig:stacking_flux_limit}
\end{figure}

The upper limit of emission additional to the alert+ events is $\Phi_{90\%, 100\rm{TeV}}^{\nu_{\mu} + \bar{\nu}_{\mu}, \rm{stacked}} = 4.2 \times 10^{-16}~(\rm{TeV}~\rm{cm}^2~\rm{s})^{-1}$ for a spectral index of $\gamma=2$ and within the energy range from 4.2~TeV to 3.6~PeV. For comparison, the diffuse astrophysical neutrino flux is $\Phi_{\text{diffuse},100\rm{TeV}} = 1.44 \times 10^{-15}~(\rm{TeV}~\rm{cm}^2~\rm{s}~\rm{sr})^{-1} $ in the range of 15~TeV to 5~PeV \citep{2022ApJ...928...50A} with a spectral index of $\gamma = 2.37$. Integrating over the energy range where both the diffuse flux and $\Phi_{90\%, 100\rm{TeV}}^{\nu_{\mu} + \bar{\nu}_{\mu}, \rm{stacked}}$ overlap, $\Phi_{90\%, 100\rm{TeV}}^{\nu_{\mu} + \bar{\nu}_{\mu}, \rm{stacked}}$ corresponds to 1.6\% of the astrophysical diffuse flux. To constrain the maximal possible emission from the alert+ regions, including the highest-energy events, we include the alert+ events just for the following limit. Thus, considering the total emission of all 122 regions, including alert+ events, we get an upper flux limit of $\Phi_{90\%,100\rm{TeV}}^{\nu_{\mu} + \bar{\nu}_{\mu}, \rm{with~alerts}}= 1.2 \times 10^{-15}~(\rm{TeV}~\rm{cm}^2~\rm{s})^{-1}$ for the energy range from 4.2~TeV to 3.6~PeV, which corresponds to 4.5\% of the diffuse astrophysical neutrino flux where both fluxes overlap in their energy range (see Figure \ref{fig:stacking_flux_limit}). We repeat the upper flux limit calculation with the same spectral index as for the astrophysical diffuse flux and get a limit of $ 2.1 \times 10^{-16}~(\rm{TeV}~\rm{cm}^2~\rm{s})^{-1}$ (1.5\% of the astrophysical diffuse flux) excluding alert+ events, and $1.1 \times 10^{-15}~(\rm{TeV}~\rm{cm}^2~\rm{s})^{-1}$ including alert+ events (8\% of the astrophysical diffuse flux) at 100~TeV. For $\gamma = 2.37$, the energies of the simulated detected events range from 0.6~TeV to 1~PeV. This energy range differs from the previous range for $\gamma=2$. The energy distribution of the signal events depends on the simulated energy spectral index. There are more neutrinos in lower energies if the simulated energy spectrum is softer compared to a harder emission.

The lack of lower energy neutrino emission (compared to IceCube alert+ events) could be caused by various scenarios. It is, for example, possible that some sources flare in neutrinos, emitting mainly high-energy neutrinos. Another possibility might be a hard neutrino emission, i.e., $\gamma \leq 1$ (for example, models proposed in \citet{Waxman:1998yy, 2022MNRAS.510.2671P}). The atmospheric background would dominate the lower-energy neutrino emission. The higher-energy neutrino emission would be detected as single high-energy events, given IceCube's effective area \citep{2020time_integrated_neutrino_sources, 2021arXiv210109836I}. This matches our observation. However, there are many different scenarios that agree with this work. In these cases, different source populations or states produce different neutrino spectra compared to one continuous power law. Another possible scenario including a source population emitting single power-laws is described in \citet{Abbasi_2023_alert_population}. Our result agrees with the high-density scenario presented in Section 6. There, a high-density source population with low individual fluxes (with an $E^{-2.5}$ energy spectrum) is the origin of alert events. Due to the sheer number of sources, we would be able to detect flux fluctuations in high-energies as alert events without a detectable lower-energy component. In lower energies, the flux would be too low to be detected and it would require a simultaneous fluctuation in both lower and higher energies such that both components could be detected from the same object. 

\subsection{Transient sources}
\begin{figure}
    \centering
    \includegraphics[width=0.9\textwidth]{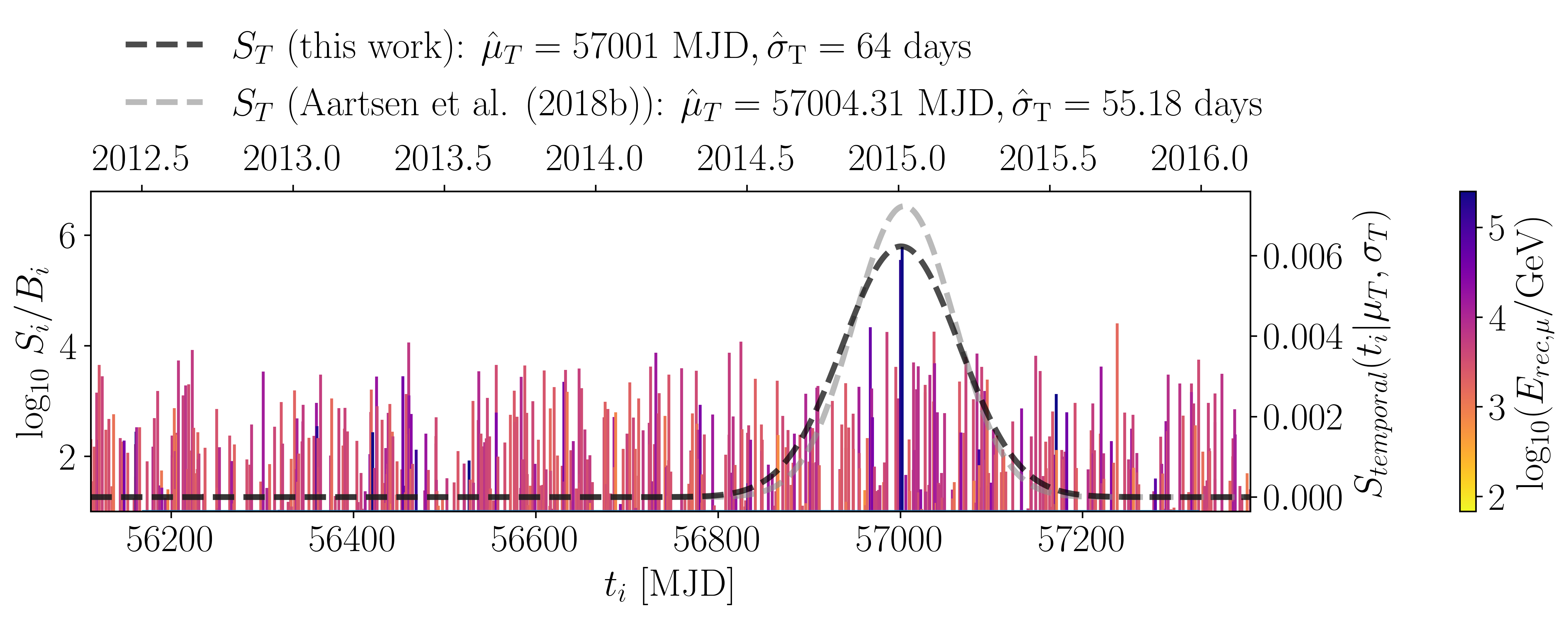}
    \caption{Logarithm of the signal-over-background ratio, $\log_{10} S_i/B_i$, distribution of individual events, $i$, versus their detection time, $t_i$, between 2012 and 2016. The $\log_{10} S_i/B_i$ values are for the best-fit position (close to TXS~0506+056) and the best-fit spectral index. The color indicates the reconstructed muon energy, $E_{\text{rec}, \mu}$, increasing from light to dark. The black-dashed line shows this work's best-fit time pdf $S_{T}$ (with the y-axis on the right). It agrees with the grey-dashed pdf of  \citet{2018Sci...361..147I}. }
    \label{fig:txs_sob}
\end{figure}

In our search for transient sources, we look for the most significant transient neutrino emission. Out of all the investigated 122 alert+ origins, the most significant transient emission is the neutrino flare with the seed alert IceCube-170922A, which is associated with the blazar TXS~0506+056. Our search yields a local p-value of  $0.14\%$ (or a significance of $3\sigma$). The main differences between the search in \citet{2018Sci...361..147I} and this work are: 
\begin{itemize}
    \item We have no external trigger in this work, whereas \citet{2018Sci...361..147I} was triggered by the observation of a flaring blazar.
    \item We use 11 years of recalibrated IceCube muon data, improving directional and energy reconstruction. For a discussion of how the contributing events are affected, see Appendix \ref{appendix:transient_source_analysis}.
    \item We include a fit for the best source position and use expectation maximization to identify the time of the neutrino flare.
\end{itemize} 
The corresponding flare is centered around a mean flare time $\hat{\mu}_T = 57001_{-26}^{+38}$~MJD and has a width of $\hat{\sigma}_T=64_{-10}^{+35}$~days. These properties agree with \citet{2018Sci...361..147I}, as shown in Figure \ref{fig:txs_sob}. When correcting for the look-elsewhere effect, the global p-value is $p_{\text{global}} = 0.156$, which is not significant. Such a trial correction does not apply for the search reported in \citet{2018Sci...361..147I}. Table \ref{tab:results_timedep} lists all results for the investigated regions. 

The best-fit parameter can yield insight into the source emission. However, as mentioned in Section \ref{sec:method} and Appendix \ref{appendix:bias}, the best-fit results and the resulting flux estimations are biased. The best-fit result of the number of neutrinos in the neutrino flare is $\hat{n}_S = 12^{+9}_{-7}$ with a spectral index of $\hat{\gamma} = 2.3 \pm 0.4$. This corresponds to an average flux of $\Phi_{100\rm{TeV}}^{\nu_{\mu} + \bar{\nu}_{\mu}} = 1.1^{+0.9}_{-0.8} \times 10^{-15}~(\rm{TeV}~\rm{cm}^2~\rm{s})^{-1}$ in the energy range of 3.5~TeV to 213~TeV during the period of the neutrino flare. The corresponding single flavor neutrino and anti-neutrino fluence, the flux integrated over the flaring period ($\hat{\mu}_T - 2\hat{\sigma}_T$ to $\hat{\mu}_T + 2\hat{\sigma}_T$), is $J_{100\rm{TeV}}^{\nu_{\mu} + \bar{\nu}_{\mu}} = 1.2 ^{+1.0} _{-0.8} \times 10^{-8} $~(TeV~cm$^2$)$^{-1}$. This flux estimation also agrees with \citet{2018Sci...361..147I}, as shows the all-flavor neutrino flux (three times $\Phi_{100\rm{TeV}}^{\nu_{\mu} + \bar{\nu}_{\mu}}$) in Figure \ref{fig:txs_sed}. In Appendix \ref{appendix:transient_source_analysis}, we compare the events contributing to the neutrino flare in this analysis and previous works and explain why the errors differ.

\begin{figure}
\centering
    \includegraphics[width=0.8\textwidth]{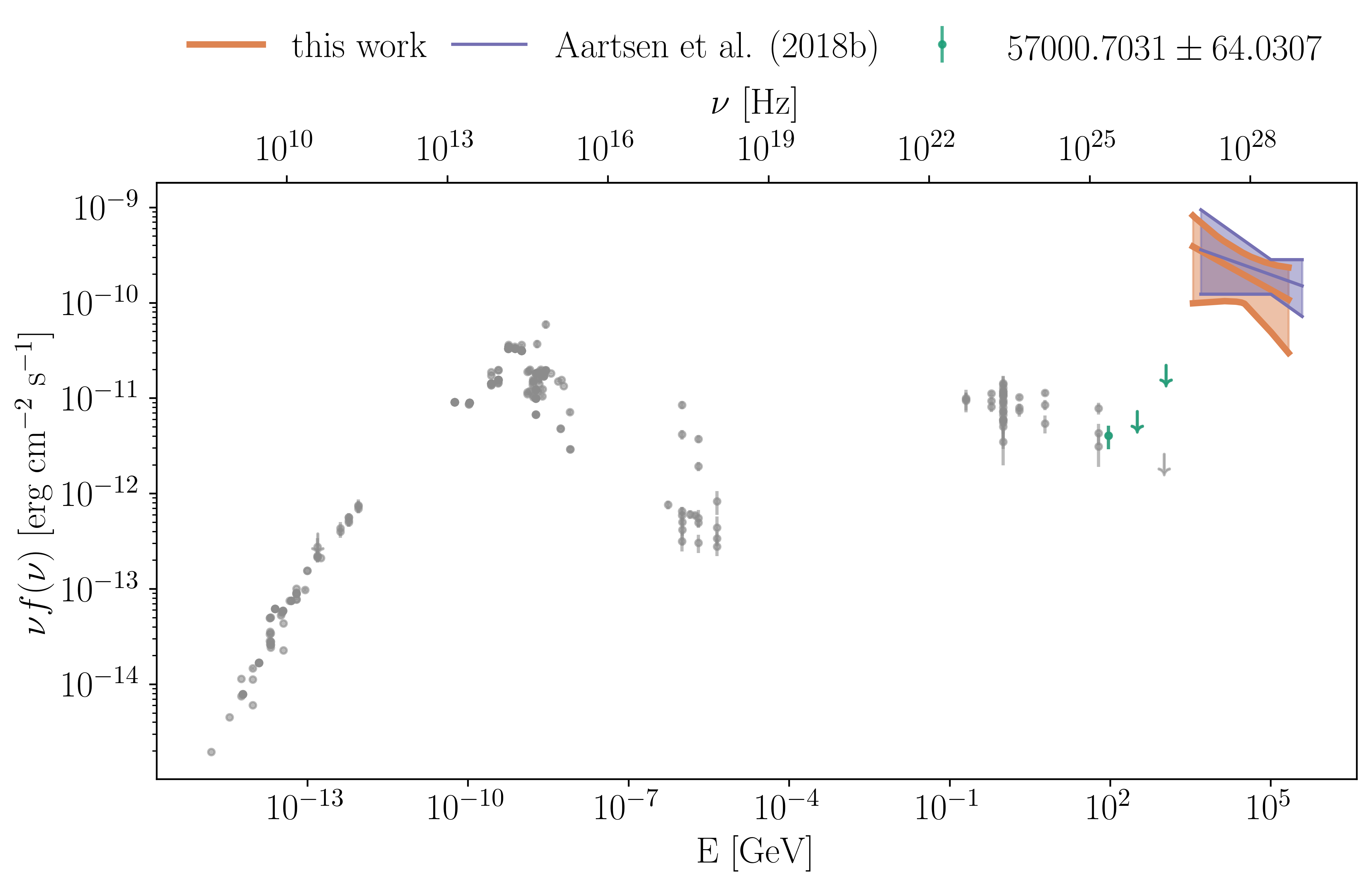}
    \caption{Spectral energy distribution of TXS~0506+056 in photons (grey dots) and neutrinos during the time of the neutrino flare (bands). The green dots (arrows) show gamma-ray emission (upper limits) during the time window of the neutrino flare detected by Fermi-LAT \citep{Ackermann_2012}. This work's all-flavor neutrino flux during the flare (orange band, $3 \times \Phi_{100\text{TeV}}^{\nu_{\mu} + \bar{\nu}_{\mu}}$) agrees with the all-flavor flux given in \citet{2018Sci...361..147I} (dark purple band). Data for the photon SED from \citet{2003MNRAS.341....1M, 2007ApJS..171...61H, 2007MNRAS.376..371J, 2007AJ....133.1947N, 1998AJ....115.1693C, 1994ApJS...91..111W, 2011A&A...536A...7P, 1996ApJS..103..427G, 1992ApJS...79..331W, 2014A&A...571A..28P, 2015arXiv150702058P, 2010AJ....140.1868W, 2011MNRAS.411.2770B, 2013A&A...551A.142D, 0067-0049-210-1-8, 1999A&A...349..389V, 2016A&A...588A.103B, 2010ApJS..188..405A, 2012ApJS..199...31N, 2015ApJS..218...23A, argoGammaRay, 2018arXiv180508505G}.}
    \label{fig:txs_sed}
\end{figure}

For transient emission, the lack of additional lower-energy neutrino emission (besides the reported local evidence associated with TXS~0506+056) can imply various scenarios. One is that neutrino flares occur rarely or might not necessarily be connected to the production sites of high-energy neutrinos. Similarly to Section \ref{sec:results_continuous_sources}, it could also indicate that these neutrino sources emit a very hard energy spectrum, for example, with $\gamma \leq 1$.

\section{Conclusion}

Our study focused on the origin of IceCube's highest energy events, or alert+ events, to identify potential sources of additional neutrino emission. To achieve this, we systematically scanned the 90\% uncertainty contours of reconstructed alert+ events, with a resolution of 0.2 degrees, to determine the most significant source position. We assumed that the emission followed a power-law distribution, $\propto E^{-\gamma}$, with $\gamma$ ranging from 1.5 to 4.

Our analysis found no evidence for continuous emission from a single source, as the data were consistent with the background assumption. Therefore, we placed a constraint on the overall combined flux from all positions, which was found to be 1.6\% of the diffuse astrophysical neutrino flux observed by IceCube (for $\gamma = 2$). If we included the alert+ events in the analysis, we could constrain all expected emissions from their respective directions to no more than 4.5\% of the diffuse astrophysical neutrino flux (for $\gamma = 2$). For a source spectral index similar to the diffuse astrophysical neutrino flux ($\gamma = 2.37$), we constrain the overall combined flux to be less than 1.5\% (excluding the alert+ events) and less than 8\% (including the alert+ events) of the diffuse astrophysical neutrino flux. This indicates that different source populations or states produce different neutrino spectra compared to one continuous power law. 
 
Our investigation confirmed the neutrino flare associated with the blazar TXS~0506+056 as the most significant transient emission from all investigated positions, with a local significance of about $3 \sigma$. When we corrected for the look-elsewhere effect in this analysis, the global significance was 15.6\%, consistent with the background expectation. The parameters of the neutrino flare in this study using recalibrated data agreed with previously published results. 
We identified a Gaussian time window with a center at $57001 ^{+38}_{-26}$~MJD, and a width of $64^{+35}_{-10}$ days as the best fit and estimated that $12^{+9}_{-6}$ neutrinos were detected during the flare with a best-fit spectral index of $\hat{\gamma} = 2.3 \pm 0.4$. This corresponds to a single flavor neutrino fluence of $J_{100\rm{TeV}}^{\nu_{\mu} + \bar{\nu}_{\mu}} = 1.2 ^{+1.0}_{-0.8} \times 10^{-8}$~(TeV~cm$^2$)$^{-1}$ and an average flux of $\Phi_{100\rm{TeV}}^{\nu_{\mu} + \bar{\nu}_{\mu}} = 1.1^{+0.9}_{-0.8} \times 10^{-15}$~(TeV~cm$^2$~s)$^{-1}$ during the $2\sigma_T$ time window.
However, we find no other alert+ event with a similar local significance. TXS~0506+056 remains the only source candidate where we find a connection of a high-energy alert and a lower energetic neutrino emission.

For neither continuous nor transient emission did we find evidence of a lower energy neutrino component. This can be explained in various scenarios. One is a hard neutrino spectrum with $\gamma \leq 1$. In such a scenario, atmospheric background noise would dominate the lower energy range, while the higher energy range would yield single high-energy events. It could also be caused by a high-density source population as investigated in \citet{Abbasi_2023_alert_population}, where high-energy events are the result of fluctuations from a large population of sources with individually weak fluxes. In this case, the lower energy flux would still be too low to be detected. Our finding also suggests neutrino flares may be rare or produced at different sites than IceCube alert+ events or that there are sources mainly emitting high-energy neutrinos.

\section*{Acknowledgements}
The IceCube Collaboration acknowledges significant contributions to this manuscript from Martina Karl. The authors gratefully acknowledge the support from the following agencies and institutions: USA {\textendash} U.S. National Science Foundation-Office of Polar Programs,
U.S. National Science Foundation-Physics Division,
U.S. National Science Foundation-EPSCoR,
U.S. National Science Foundation-Office of Advanced Cyberinfrastructure,
Wisconsin Alumni Research Foundation,
Center for High Throughput Computing (CHTC) at the University of Wisconsin{\textendash}Madison,
Open Science Grid (OSG),
Partnership to Advance Throughput Computing (PATh),
Advanced Cyberinfrastructure Coordination Ecosystem: Services {\&} Support (ACCESS),
Frontera computing project at the Texas Advanced Computing Center,
U.S. Department of Energy-National Energy Research Scientific Computing Center,
Particle astrophysics research computing center at the University of Maryland,
Institute for Cyber-Enabled Research at Michigan State University,
Astroparticle physics computational facility at Marquette University,
NVIDIA Corporation,
and Google Cloud Platform;
Belgium {\textendash} Funds for Scientific Research (FRS-FNRS and FWO),
FWO Odysseus and Big Science programmes,
and Belgian Federal Science Policy Office (Belspo);
Germany {\textendash} Bundesministerium f{\"u}r Bildung und Forschung (BMBF),
Deutsche Forschungsgemeinschaft (DFG),
Helmholtz Alliance for Astroparticle Physics (HAP),
Initiative and Networking Fund of the Helmholtz Association,
Deutsches Elektronen Synchrotron (DESY),
and High Performance Computing cluster of the RWTH Aachen;
Sweden {\textendash} Swedish Research Council,
Swedish Polar Research Secretariat,
Swedish National Infrastructure for Computing (SNIC),
and Knut and Alice Wallenberg Foundation;
European Union {\textendash} EGI Advanced Computing for research;
Australia {\textendash} Australian Research Council;
Canada {\textendash} Natural Sciences and Engineering Research Council of Canada,
Calcul Qu{\'e}bec, Compute Ontario, Canada Foundation for Innovation, WestGrid, and Digital Research Alliance of Canada;
Denmark {\textendash} Villum Fonden, Carlsberg Foundation, and European Commission;
New Zealand {\textendash} Marsden Fund;
Japan {\textendash} Japan Society for Promotion of Science (JSPS)
and Institute for Global Prominent Research (IGPR) of Chiba University;
Korea {\textendash} National Research Foundation of Korea (NRF);
Switzerland {\textendash} Swiss National Science Foundation (SNSF).

%

\vspace{5mm}




 
\appendix

\section{Parameter recovery}\label{appendix:bias}
When testing the method, as described in Section \ref{sec:method}, with Monte Carlo simulations \citep{mythesis}, the best-fit number of signal neutrinos, $n_S$, and source spectral index, $\gamma$, show a bias compared to the true simulated source properties. For sources with simulated hard spectral indices (i.e., $\gamma = 2$), there is a tendency to fit slightly softer spectra and a slightly larger number of signal neutrinos. For example, simulating an average of 10 neutrinos with $\gamma = 2$ results in a mean best-fit of $\hat{n}_S = 16$ and $\hat{\gamma} = 2.25$. For simulated sources following softer spectral indices (i.e., $\gamma = 3$), the tendency is reversed to fitting slightly harder spectral indices and smaller numbers of signal neutrinos. 

Several aspects influence this bias. One is a simplified spatial distribution in the form of a Rayleigh distribution (see Equation (\ref{eq:signal_timeint_gen})). This is corrected using a kernel density estimation (KDE) approach, for example, in \cite{doi:10.1126/science.abg3395}. However, the KDE approach is, so far, only feasible in the northern sky. Since we search for neutrino sources also from the southern sky, we chose the simplified method. Another aspect is that weak sources emitting only few neutrinos are not always found doing the position scan since background fluctuations can dominate these weak sources. For example, for a continuous emission over 11 years, the mean distance between the best-fit source position and the actual simulated source is smaller than 0.3 degrees for a flux resulting in 5 signal neutrinos on average. This also means that the best-fit $n_S$ will be larger than 0 in many cases with no neutrino source since the algorithm will find the position with the largest background fluctuation. Hence, correcting this bias is not straightforward, and this analysis is mainly sensitive to strong neutrino sources.

For transient sources, the bias is smaller. In the same example as above, 10 neutrinos with $\gamma = 2$ emitted over a period of $\sigma_T \approx 55$~days are a much stronger signal compared to 10 neutrinos over 11 years. Hence in this specific case, the mean best-fit $\hat{n}_S = 12$ and the best-fit $\hat{\gamma} = 2.1$. However, we still face the case that background fluctuations can dominate weak neutrino emission (in the case of $\sigma_T \approx 55$~days, anything below 5 neutrinos is difficult), which makes correcting this bias challenging. We have decided not to include an, at best incomplete, correction in this work. For now, measurements of point-source fluxes are only possible with the KDE approach.

\section{Transient sources analysis}\label{appendix:transient_source_analysis}

\begin{figure}[h!]
    \begin{minipage}{0.5\textwidth}
    \includegraphics[width=\textwidth]{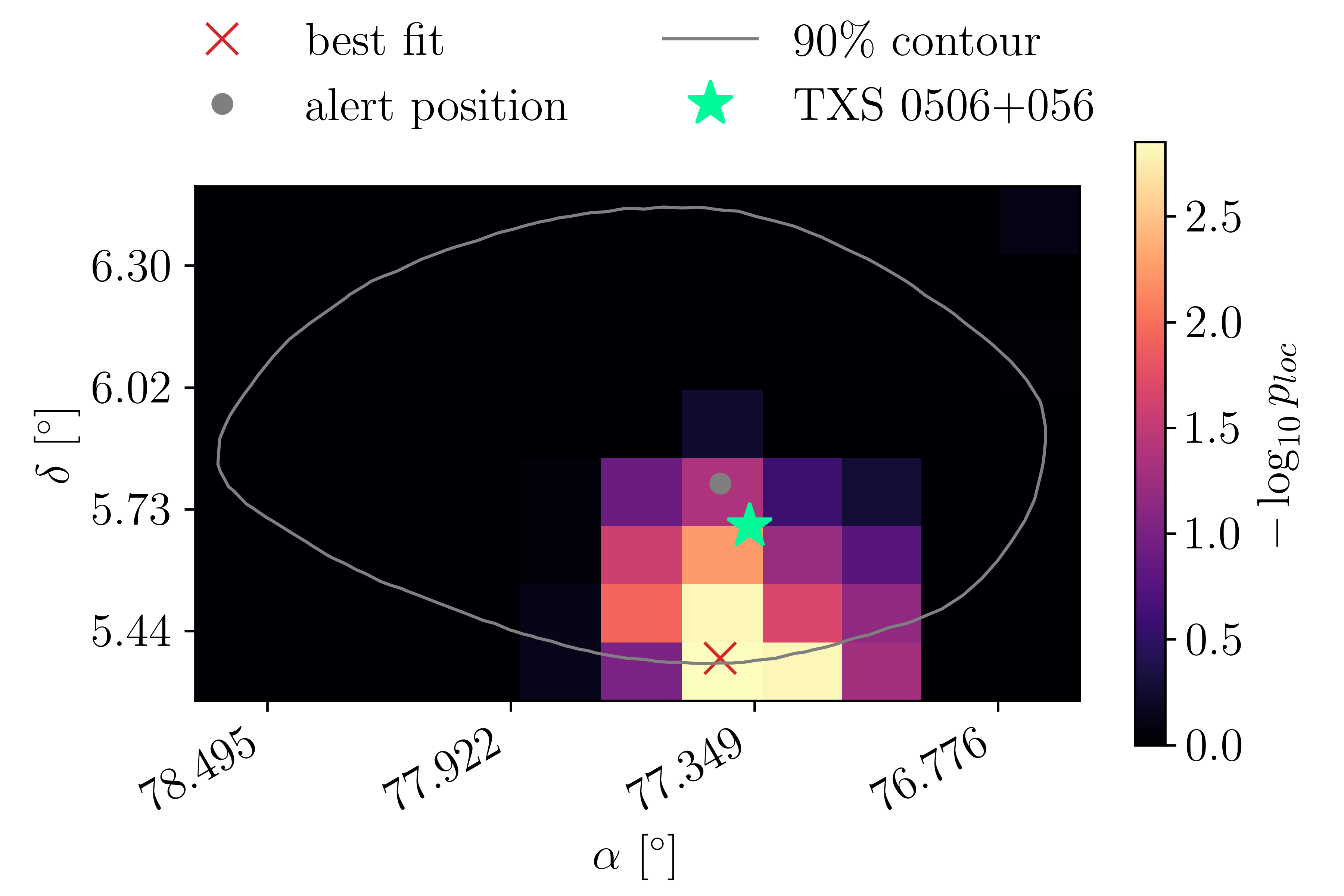}
    \end{minipage}
    \begin{minipage}{0.5\textwidth}
    \centering
    \includegraphics[width=0.8\textwidth]{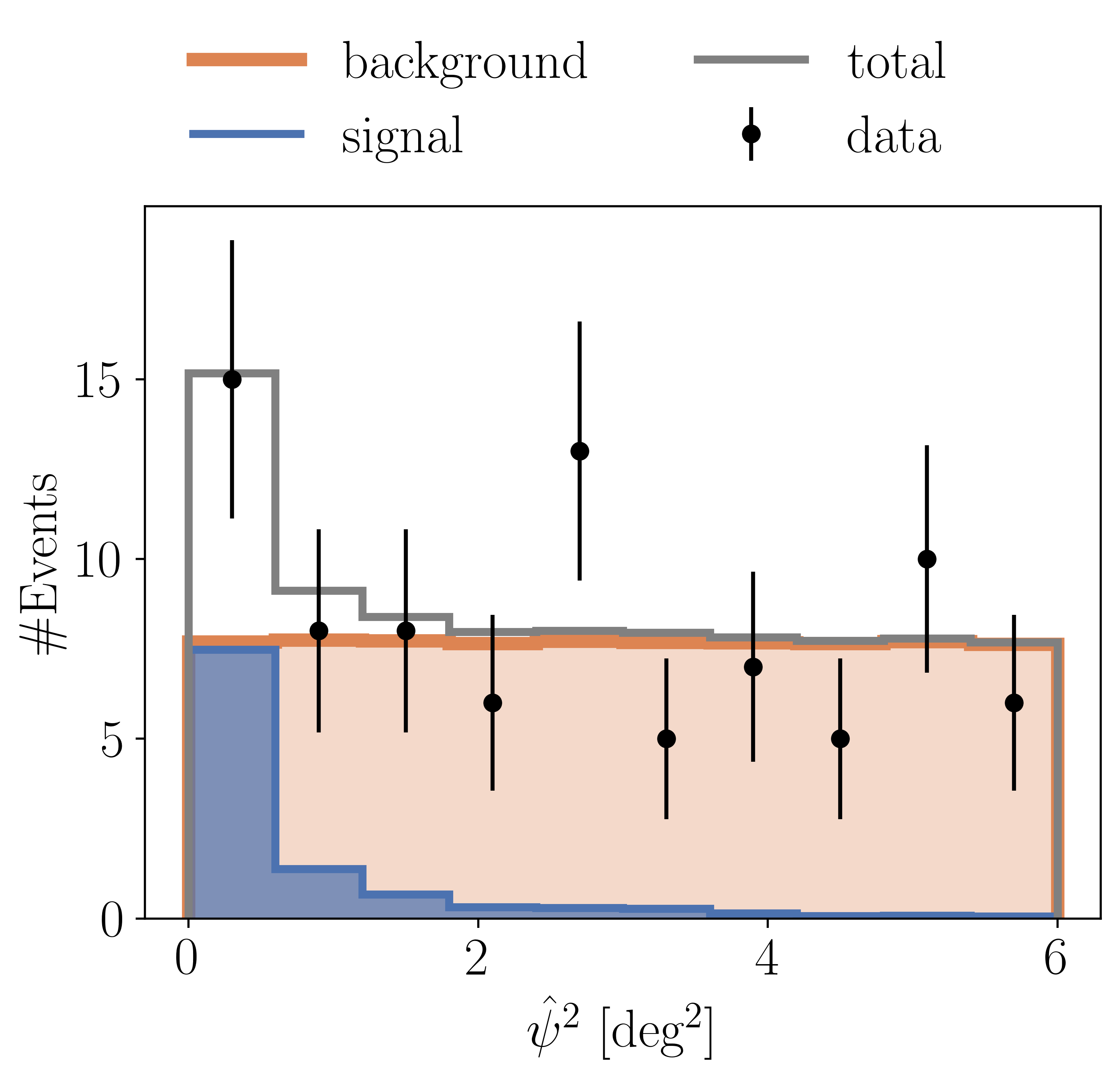}
    \end{minipage}
    \caption{\textbf{Left:} P-value map of the alert region of IceCube-170922A. The grey dot indicates the reconstructed direction of IceCube-170922A, and the grey contour shows the 90\% uncertainties of the reconstruction. The red cross marks the best-fit position of the position scan ($0.6^{\circ}$ from the reconstructed alert position). The star shows the location of TXS~0506+056. All black bins have p-values close to 1. \textbf{Right:} Number of events at binned squared angular distances, $\hat{\Psi}^2$, between TXS~0506+056 and the reconstructed event directions during the neutrino flare ($57001~\rm{MJD} \pm 2 \times 64$~days). Scrambled data in right ascension provides the background (blue), and Monte Carlo simulations for the best-fitted flux ($n_S = 12$ and $\gamma = 2.31$) yield the signal (orange). The grey line combines the background with the signal and matches the data points (black). The data are shown with 68\% uncertainties.}
    \label{fig:txs_ang_dist}
\end{figure}

Figure \ref{fig:txs_ang_dist} shows the p-value map of the scanned region around IceCube-170922A on the left. The most significant position is within $0.5^{\circ}$ from TXS~0506+056. The right panel of Figure \ref{fig:txs_ang_dist} shows a histogram of the angular distance of events from TXS~0506+056. There is a clustering of events around the source position. The signal events for this plot are simulated according to the best-fit result of the likelihood ratio test ($\hat{n}_S = 12$, $\hat{\gamma} = 2.3$). The background distribution is scrambled data in right ascension. The signal on top of the background flux matches the observed data. 

\begin{figure}[h!]
    \begin{minipage}{0.33\textwidth}
        \includegraphics[width=\textwidth]{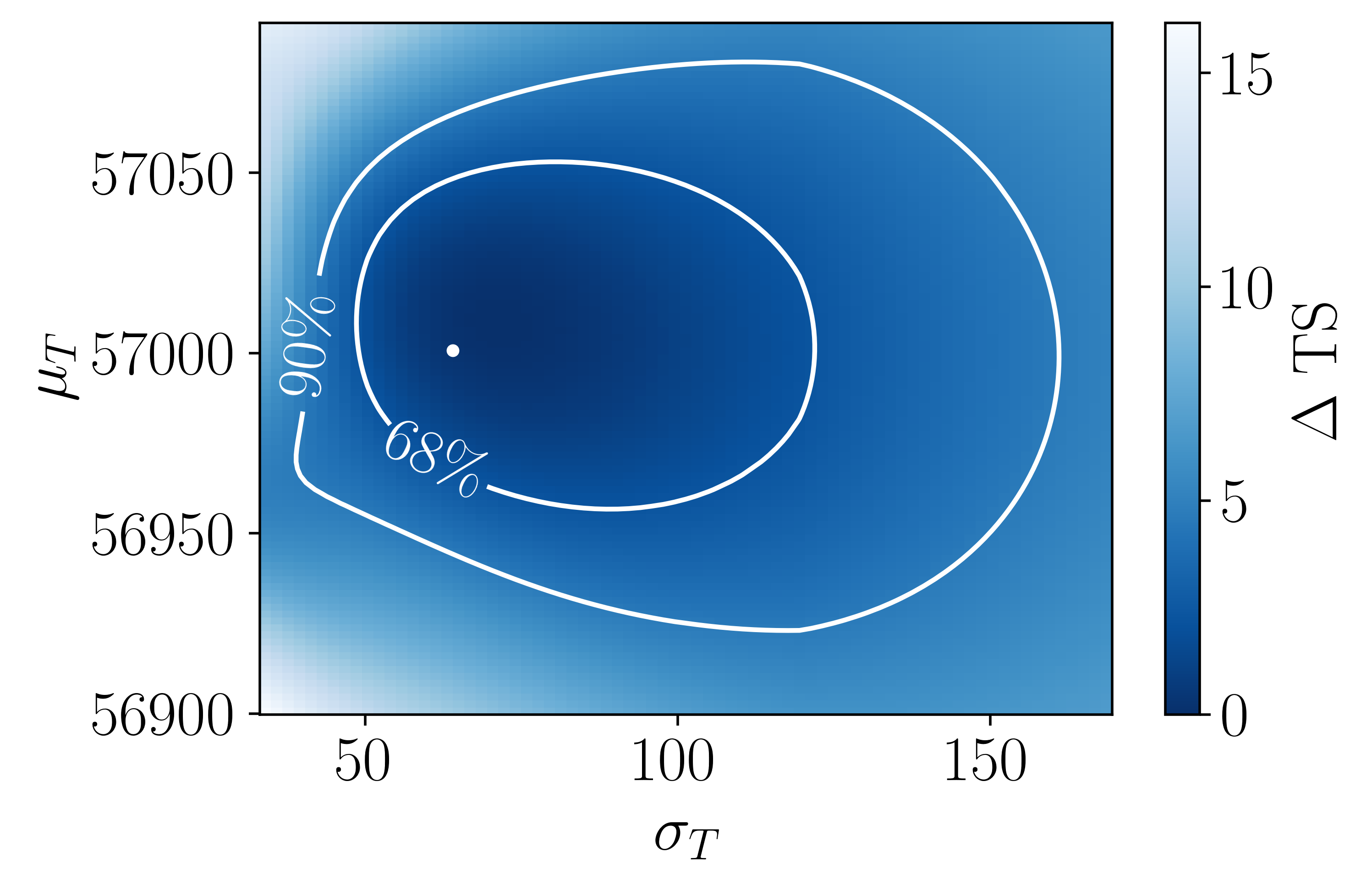}
    \end{minipage}
        \begin{minipage}{0.33\textwidth}
        \includegraphics[width=\textwidth]{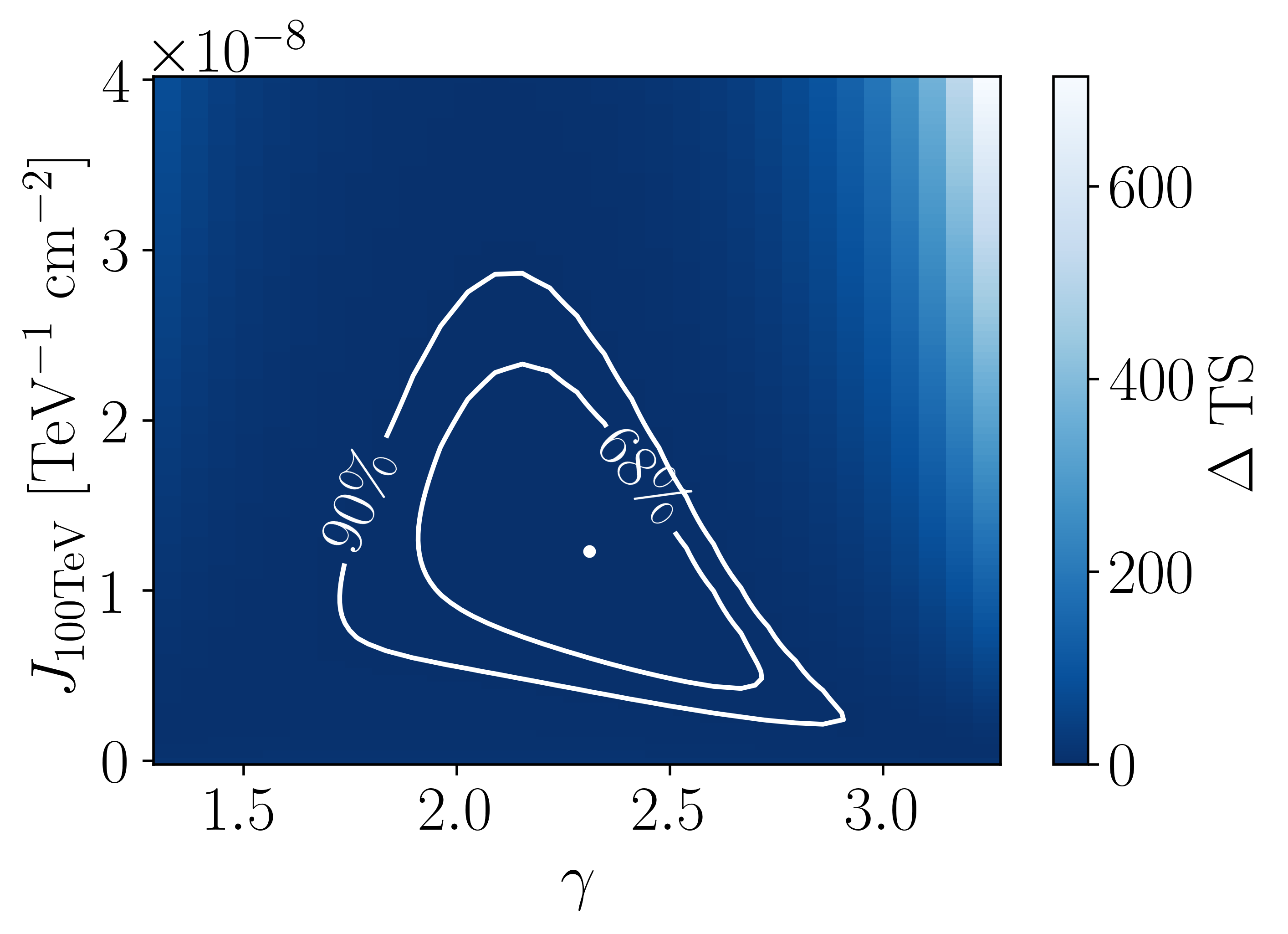}
    \end{minipage}
        \begin{minipage}{0.33\textwidth}
        \includegraphics[width=\textwidth]{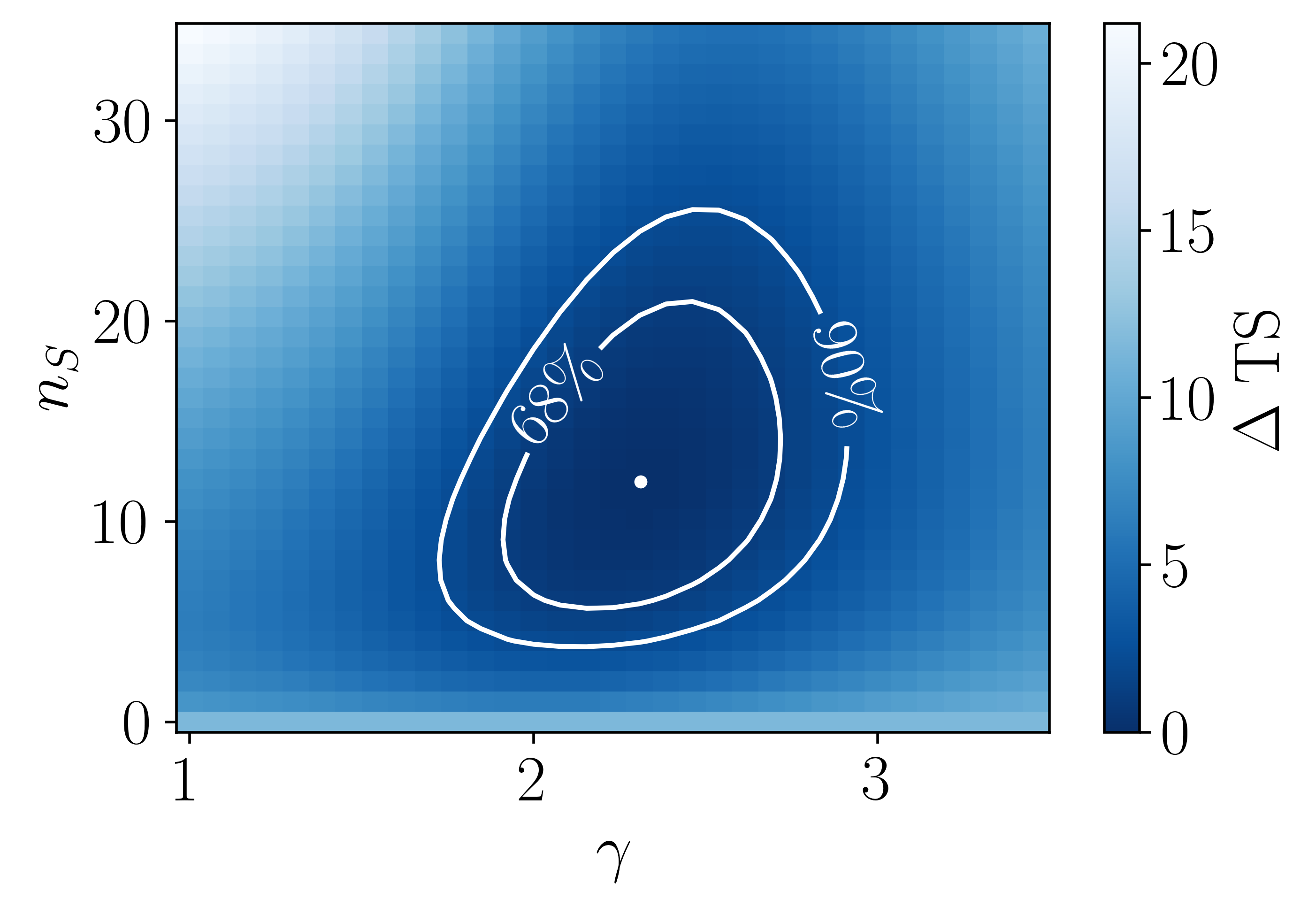}
    \end{minipage}
    \caption{Change of the test statistic value for the different likelihood parameters. \textbf{Left:} Profiled change for different $\mu_T$ and $\sigma_T$. $n_S$ and $\gamma$ are optimized at each step. The 68\% uncertainties are $\hat{\mu}_T = 57001 ^{+38}_{-26}~\text{MJD}$ and $\hat{\sigma}_T = 64^{+35}_{-10}$~days. \textbf{Center:} Change when varying the signal fluence $J_{100\rm{TeV}}^{\nu_{\mu} + \bar{\nu}_{\mu}}(n_S, \gamma)$. The 68\% uncertainties on the fluence are $J_{100\rm{TeV}}^{\nu_{\mu} + \bar{\nu}_{\mu}} = 1.2^{+1.0}_{-0.8}\times 10^{-8}~(\rm{TeV}~\rm{cm}^2)^{-1}$. \textbf{Right:} Variation when changing $n_S$ and $\gamma$. The 68\% uncertainties are $\hat{n}_S = 12 ^{+9}_{-7}$ and $\hat{\gamma} = 2.3 \pm 0.4$.}
    \label{fig:timedep_results_contours}
\end{figure}

To determine the uncertainties of the best-fit values, we run a likelihood scan over the parameter space and use Wilk's theorem \citep{10.1214/aoms/1177732360} to determine the 68\% and 90\% contours (see Figure \ref{fig:timedep_results_contours}). These contours are relevant for the two-dimensional uncertainties of the flux as in Figure \ref{fig:txs_sed}. For the time, we determine the profiled change of the test statistic for different $\mu_T$ and $\sigma_T$. The best $n_S$ and $\gamma$ are fitted for each value. The 68\% uncertainties determined by a profiled change of the test statistic are $\hat{\mu}_T = 57001 ^{+38} _{-26}~\text{MJD}$ and $\hat{\sigma}_T = 64 ^{+35} _{-10}$~days. The one-dimensional errors for fluence, number of signal neutrinos, $n_S$, and spectral index $\gamma$, are determined with the profiled change of the test statistic where the mean flaring time and the flare width are kept fixed to the best-fit values. For the signal fluence the 68\% uncertainties are are $J_{100\rm{TeV}}^{\nu_{\mu} + \bar{\nu}_{\mu}} = 1.2^{+1.0}_{-0.8}\times 10^{-8}~(\rm{TeV}~\rm{cm}^2)^{-1}$ and for $n_S$ and $\gamma$ we get $\hat{n}_S = 12 ^{+9}_{-7}$ and $\hat{\gamma} = 2.3 \pm 0.4$.

 Table \ref{tab:contributing_events} lists the top 14 contributing events to the neutrino flare, sorted by their $S_i/B_i$ value multiplied by $S_{T}$. We compare this with a previous data sample \citep{2021arXiv210109836I} (for events also included in that sample) to emphasize how the updated photomultiplier calibration affects the reconstructed direction, angular error, and energy. 

\begin{table}
    \caption{\textbf{Left:} Top 14 events with the strongest contribution to the neutrino flare of TXS~0506+056, sorted by significance. \textbf{Right:} The same events in the data sample published in \cite{2021arXiv210109836I}. The last row states the ranking of the contribution in previous analyses. The data set used in this work has improved directional and energy reconstruction. Some events have shifted in position and have slightly different energies.}
       \begin{center}

\begin{tabular}{c | c c c c | c c c c c }
     \tableline
    & \multicolumn{4}{c}{This work} & \multicolumn{5}{c}{\cite{2021arXiv210109836I}} \\
    MJD & RA  & Dec & $\sigma$    & $\log_{10}(E/\rm{GeV})$&RA  & Dec & $\sigma$   &$\log_{10}(E/\rm{GeV})$ & Ranking\\
        & (deg) & (deg) & (deg) & &(deg) & (deg) & (deg) && \\
     \tableline
   56940.9084  &  77.36  &  5.42  &  0.20  &  3.81 &  77.35    &5.42    &0.20    &3.97 & 1\\
   57009.5301  &  77.36  &  5.53  &  0.34  &  3.85 &  77.32    &5.50    &0.34    &3.91 & 2\\
   56973.3971  &  77.03  &  5.01  &  0.39  &  3.61 &  77.05    &5.05    &0.40    &3.71 & 12 \\
   57112.6530  &  77.39  &  5.32  &  0.20  &  3.23 &  77.43    &5.34    &1.09    &3.46 & 7 \\
   57072.2088  &  77.13  &  5.04  &  0.42  &  3.50 &  76.35    &5.22    &0.36    &3.43 & 9 \\
   56981.1313  &  76.20  &  6.13  &  0.63  &  4.03 &  76.16    &6.19    &0.43    &4.13 & 5 \\
   57089.4395  &  77.67  &  5.91  &  0.20  &  3.62 &  77.71    &5.90    &0.20    &3.69 & 3 \\
   56927.8601  &  77.43  &  4.93  &  0.39  &  3.46 &  77.39    &4.93    &0.33    &3.53 & 13 \\
   56955.7917  &  77.61  &  5.58  &  0.51  &  2.99 &  77.60    &5.56    &0.48    &3.09 & 6 \\
   57072.9895  &  76.05  &  6.80  &  1.97  &  4.09 &  76.35    &5.22    &0.36    &4.17 & 4 \\
   56940.5215  &  77.82  &  5.79  &  0.44  &  2.80 & -- & -- & -- & -- & -- \\
   57031.8224  &  77.64  &  4.61  &  0.76  &  2.96 & -- & -- & -- & -- & -- \\
   56937.8189  &  77.77  &  6.29  &  0.63  &  2.98 &  77.75    &6.23    &0.63    &2.91 & 11 \\
   56983.2476  &  77.47  &  6.80  &  0.92  &  3.09 & -- & -- & -- & -- & -- \\
   \tableline
\end{tabular}
\label{tab:contributing_events}
\end{center}
\end{table}

\begin{figure}
    \centering
    \includegraphics[width=0.8\textwidth]{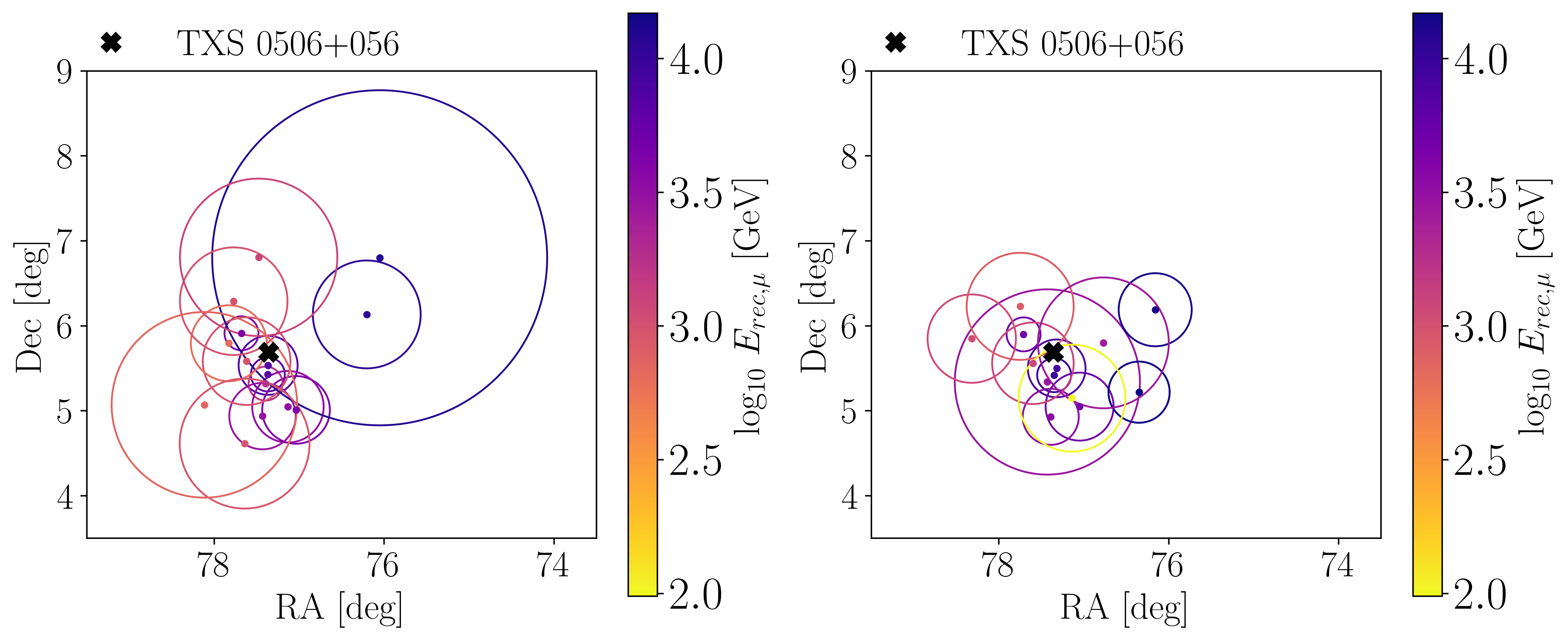}
    \caption{Position and energy (color) of the 14 most contributing events to the TXS~0506+056 neutrino flare. The circles show the uncertainty of the directional reconstruction, $\sigma_i$. \textbf{Left:} The 14 most contributing events from the data sample used in this work (see Table \ref{tab:contributing_events}). \textbf{Right:} The 14 most contributing events from the old data sample \citep{2021arXiv210109836I}.}
    \label{fig:contributing_events_pos}  
\end{figure}

The improved directional and energy reconstruction has changed the contributing events compared to previous analyses \citep{2018Sci...361..147I, 2021arXiv210109836I}. Most of the significance is caused by the two most contributing events, which remain the same (see also \citet{mythesis}). However, their position is shifted, and their energy is changed. For the remaining events, the contributing order has changed, or the events themselves differ. Figure \ref{fig:contributing_events_pos} shows the position and energy of the 14 most contributing events to the neutrino flare from the previous data set (left) and the improved data used in this work (right). The event with the largest error region ($\sigma = 1.9^{\circ}$) on the right panel is also included in the left panel. However, the uncertainty was underestimated in the previous data sample ($\sigma = 0.36^{\circ}$), and its position has shifted. 

\FloatBarrier
\newpage
\section{IceCube alert+ events}

\begin{center}
    \begin{longtable}{l l l c c } 
        \caption[Table of all IceCube alert+ events analyzed in this work]{All alert events (track name starting with ``IC'') and high-energy tracks (track name starting with ``DIF'' (selected from \citet{2022ApJ...928...50A}), ``EHE'' (extremely-high-energy), or ``HESE'' (high-energy-starting event)) investigated in this work. The track name includes the time of detection in the format yymmdd. In the case of alert events, the letter ``A'' or ``B'' is used to distinguish events detected on the same day. The time is the detection time in MJD, and R.A. and Dec list the best reconstruction coordinates with 90\% confidence level uncertainties.} \\

        \tableline
        Index & Track Name & Time [MJD] & R.A. [deg]& Dec [deg]   \\ 
        \tableline
     \endfirsthead

        \multicolumn{5}{c}{{\tablename\ \thetable{} -- continued from previous page}} \\
        \tableline
        Index & Track Name & Time [MJD] & R.A. [deg]& Dec [deg]   \\ 
        \tableline
        \endhead

        \multicolumn{5}{c}{{Continued on next page}} \\ 
        \endfoot

        \endlastfoot

 1 & DIF090813 & 55056.6983 & $29.51^{+0.40}_{-0.38}$ & $1.23^{+0.18}_{-0.22}$ \\ 
 2 & DIF091106 & 55141.1275 & $298.21^{+0.53}_{-0.57}$ & $11.74^{+0.32}_{-0.38}$ \\ 
 3 & DIF100608 & 55355.4872 & $344.93^{+3.39}_{-2.90}$ & $23.58^{+2.31}_{-4.13}$ \\ 
 4 & DIF100623 & 55370.7355 & $141.25^{+0.46}_{-0.45}$ & $47.80^{+0.56}_{-0.48}$ \\ 
 5 & DIF100710 & 55387.5362 & $306.96^{+2.70}_{-2.28}$ & $21.00^{+2.25}_{-1.56}$ \\ 
 6 & DIF100925 & 55464.8959 & $266.29^{+0.58}_{-0.62}$ & $13.40^{+0.52}_{-0.45}$ \\ 
 7 & DIF101009 & 55478.3806 & $331.09^{+0.56}_{-0.72}$ & $11.10^{+0.48}_{-0.58}$ \\ 
 8 & DIF101028 & 55497.3033 & $88.68^{+0.54}_{-0.55}$ & $0.46^{+0.33}_{-0.27}$ \\ 
 9 & HESE101112 & 55512.5516 & $110.56^{+0.80}_{-0.37}$ & $-0.37^{+0.48}_{-0.65}$ \\ 
 10 & DIF101113 & 55513.5995 & $285.95^{+1.29}_{-1.50}$ & $3.15^{+0.70}_{-0.63}$ \\ 
 11 & DIF110128 & 55589.5628 & $307.53^{+0.82}_{-0.81}$ & $1.19^{+0.35}_{-0.32}$ \\ 
 12 & EHE110304 & 55624.9548 & $116.37^{+0.73}_{-0.73}$ & $-10.72^{+0.57}_{-0.65}$ \\ 
 13 & IC110514A & 55695.0642 & $138.47^{+6.68}_{-3.78}$ & $-1.94^{+0.97}_{-1.12}$ \\ 
 14 & DIF110521 & 55702.7666 & $235.13^{+2.70}_{-1.76}$ & $20.30^{+1.00}_{-1.43}$ \\ 
 15 & IC110610A & 55722.4261 & $272.55^{+1.67}_{-2.42}$ & $35.64^{+1.30}_{-1.05}$ \\ 
 16 & IC110714A & 55756.1130 & $68.20^{+0.31}_{-1.10}$ & $40.67^{+0.44}_{-0.44}$ \\ 
 17 & DIF110722 & 55764.2196 & $315.66^{+5.91}_{-5.35}$ & $5.29^{+4.85}_{-4.72}$ \\ 
 18 & IC110902A & 55806.0922 & $9.76^{+2.85}_{-1.32}$ & $7.59^{+0.87}_{-0.86}$ \\ 
 19 & IC110907A & 55811.7946 & $196.08^{+3.92}_{-2.68}$ & $9.40^{+1.56}_{-1.05}$ \\ 
 20 & DIF110930 & 55834.4451 & $266.48^{+2.09}_{-1.55}$ & $-4.41^{+0.59}_{-0.86}$ \\ 
 21 & DIF111201 & 55896.8575 & $222.87^{+1.95}_{-7.73}$ & $1.87^{+1.25}_{-1.18}$ \\ 
 22 & IC111216A & 55911.2769 & $36.74^{+1.80}_{-2.24}$ & $18.88^{+2.46}_{-2.82}$ \\ 
 23 & IC120301A & 55987.8069 & $237.96^{+0.53}_{-0.61}$ & $18.76^{+0.47}_{-0.51}$ \\ 
 24 & IC120515A & 56062.9590 & $198.94^{+1.71}_{-1.41}$ & $32.00^{+0.97}_{-1.09}$ \\ 
 25 & IC120523A & 56070.5743 & $171.08^{+0.66}_{-1.41}$ & $26.44^{+0.46}_{-0.37}$ \\ 
 26 & IC120807A & 56146.2071 & $330.07^{+0.84}_{-0.83}$ & $1.42^{+0.59}_{-0.45}$ \\ 
 27 & IC120916A & 56186.3053 & $182.24^{+1.36}_{-1.71}$ & $3.88^{+0.68}_{-0.82}$ \\ 
 28 & IC120922A & 56192.5493 & $70.62^{+1.49}_{-1.27}$ & $19.79^{+0.91}_{-0.71}$ \\ 
 29 & IC121011A & 56211.7709 & $205.14^{+0.66}_{-0.71}$ & $-2.28^{+0.53}_{-0.56}$ \\ 
 30 & IC121026A & 56226.5995 & $169.80^{+1.32}_{-1.40}$ & $27.91^{+0.85}_{-0.88}$ \\ 
 31 & IC130127A & 56319.2800 & $352.97^{+1.32}_{-1.01}$ & $-1.98^{+0.97}_{-0.89}$ \\ 
 32 & IC130408A & 56390.1888 & $167.83^{+2.63}_{-3.96}$ & $20.66^{+1.28}_{-0.99}$ \\ 
 33 & IC130627A & 56470.1104 & $93.74^{+1.01}_{-1.15}$ & $14.17^{+1.23}_{-1.04}$ \\ 
 34 & DIF130817 & 56521.8320 & $224.89^{+0.87}_{-1.19}$ & $-4.44^{+1.21}_{-0.94}$ \\ 
 35 & IC130907A & 56542.7931 & $130.17^{+0.48}_{-0.31}$ & $-10.54^{+0.26}_{-0.30}$ \\ 
 36 & IC131014A & 56579.9092 & $32.92^{+0.87}_{-0.71}$ & $10.28^{+0.41}_{-0.57}$ \\ 
 37 & IC131023A & 56588.5585 & $301.90^{+1.02}_{-1.05}$ & $11.61^{+1.14}_{-1.30}$ \\ 
 38 & IC131124A & 56620.1451 & $285.16^{+2.20}_{-1.54}$ & $19.47^{+1.43}_{-1.46}$ \\ 
 39 & IC131204A & 56630.4701 & $288.98^{+1.10}_{-0.83}$ & $-14.21^{+0.77}_{-1.31}$ \\ 
 40 & IC140101A & 56658.4039 & $192.26^{+2.07}_{-2.37}$ & $-2.69^{+1.01}_{-0.71}$ \\ 
 41 & IC140108A & 56665.3079 & $344.66^{+0.53}_{-0.48}$ & $1.57^{+0.37}_{-0.34}$ \\ 
 42 & IC140109A & 56666.5030 & $293.12^{+0.79}_{-1.19}$ & $33.02^{+0.45}_{-0.53}$ \\ 
 43 & IC140203A & 56691.7851 & $349.58^{+2.64}_{-2.54}$ & $-13.55^{+1.14}_{-1.74}$ \\ 
 44 & DIF140522 & 56799.9614 & $349.39^{+2.89}_{-4.12}$ & $18.05^{+1.94}_{-1.80}$ \\ 
 45 & IC140609A & 56817.6364 & $106.26^{+2.68}_{-2.15}$ & $1.31^{+1.04}_{-0.86}$ \\ 
 46 & IC140611A & 56819.2044 & $110.65^{+0.53}_{-0.61}$ & $11.45^{+0.19}_{-0.19}$ \\ 
 47 & IC140705A & 56843.6687 & $25.88^{+1.85}_{-2.98}$ & $2.54^{+1.79}_{-1.76}$ \\ 
 48 & IC140923A & 56923.7211 & $169.72^{+0.70}_{-0.84}$ & $-1.60^{+0.52}_{-0.30}$ \\ 
 49 & IC140927A & 56927.1608 & $50.89^{+3.91}_{-5.14}$ & $-0.63^{+1.49}_{-1.42}$ \\ 
 50 & IC150127A & 57049.4813 & $100.37^{+1.36}_{-1.62}$ & $4.59^{+0.79}_{-0.67}$ \\ 
 51 & IC150515A & 57157.9416 & $91.49^{+0.93}_{-0.74}$ & $12.14^{+0.53}_{-0.50}$ \\ 
 52 & IC150714A & 57217.9097 & $326.29^{+1.50}_{-1.31}$ & $26.36^{+1.89}_{-2.19}$ \\ 
 53 & IC150812B & 57246.7591 & $328.27^{+0.75}_{-0.88}$ & $6.17^{+0.48}_{-0.53}$ \\ 
 54 & IC150831A & 57265.2178 & $54.76^{+0.92}_{-0.93}$ & $34.00^{+1.14}_{-1.20}$ \\ 
 55 & IC150904A & 57269.7597 & $133.77^{+0.53}_{-0.88}$ & $28.08^{+0.51}_{-0.55}$ \\ 
 56 & IC150919A & 57284.2057 & $279.54^{+1.75}_{-2.29}$ & $30.35^{+2.18}_{-1.51}$ \\ 
 57 & IC150923A & 57288.0268 & $103.23^{+0.70}_{-1.15}$ & $3.96^{+0.60}_{-0.75}$ \\ 
 58 & IC150926A & 57291.9012 & $194.55^{+0.79}_{-1.23}$ & $-4.56^{+0.94}_{-0.63}$ \\ 
 59 & IC151017A & 57312.6757 & $197.53^{+2.47}_{-2.72}$ & $19.95^{+3.00}_{-2.29}$ \\ 
 60 & IC151114A & 57340.8735 & $76.16^{+1.36}_{-1.37}$ & $12.71^{+0.65}_{-0.72}$ \\ 
 61 & IC151122A & 57348.5316 & $262.05^{+0.87}_{-1.06}$ & $-2.24^{+0.64}_{-0.67}$ \\ 
 62 & IC160104A & 57391.4438 & $79.41^{+0.83}_{-0.75}$ & $5.00^{+0.87}_{-0.97}$ \\ 
 63 & IC160128A & 57415.1835 & $263.76^{+1.10}_{-1.80}$ & $-14.90^{+1.08}_{-1.20}$ \\ 
 64 & IC160225A & 57443.8804 & $311.87^{+2.19}_{-1.77}$ & $60.06^{+1.65}_{-1.38}$ \\ 
 65 & IC160331A & 57478.5652 & $151.22^{+0.66}_{-0.66}$ & $15.48^{+0.66}_{-0.73}$ \\ 
 66 & IC160510A & 57518.6640 & $352.88^{+1.76}_{-1.45}$ & $1.90^{+0.75}_{-0.67}$ \\ 
 67 & EHE160731 & 57600.0799 & $214.50^{+0.75}_{-0.75}$ & $-0.33^{+0.75}_{-0.75}$ \\ 
 68 & IC160806A & 57606.5150 & $122.78^{+0.88}_{-1.23}$ & $-0.71^{+0.56}_{-0.56}$ \\ 
 69 & IC160814A & 57614.9069 & $200.04^{+3.12}_{-2.68}$ & $-32.13^{+1.74}_{-1.25}$ \\ 
 70 & IC160924A & 57655.7411 & $241.13^{+4.92}_{-5.89}$ & $1.34^{+3.40}_{-2.79}$ \\ 
 71 & IC161001A & 57662.4392 & $192.57^{+2.50}_{-2.07}$ & $37.12^{+1.51}_{-2.48}$ \\ 
 72 & DIF161011 & 57672.0796 & $26.38^{+0.66}_{-0.66}$ & $9.55^{+0.66}_{-0.66}$ \\ 
 73 & IC161012A & 57673.6126 & $190.06^{+2.20}_{-4.04}$ & $-7.48^{+2.18}_{-2.99}$ \\ 
 74 & IC161117A & 57709.3320 & $78.66^{+1.85}_{-1.93}$ & $1.60^{+1.91}_{-1.79}$ \\ 
 75 & IC161210A & 57732.8380 & $46.36^{+2.38}_{-0.92}$ & $15.25^{+0.93}_{-1.08}$ \\ 
 76 & IC170105A & 57758.1419 & $309.95^{+5.01}_{-7.56}$ & $8.16^{+2.00}_{-3.34}$ \\ 
 77 & IC170321A & 57833.3141 & $98.26^{+1.32}_{-0.92}$ & $-15.06^{+1.04}_{-1.20}$ \\ 
 78 & IC170514B & 57887.3002 & $227.37^{+1.23}_{-1.10}$ & $30.65^{+1.40}_{-0.99}$ \\ 
 79 & IC170626A & 57930.5193 & $280.99^{+3.03}_{-1.63}$ & $8.80^{+1.13}_{-0.90}$ \\ 
 80 & IC170704A & 57938.2926 & $230.45^{+1.67}_{-1.71}$ & $23.36^{+1.10}_{-0.89}$ \\ 
 81 & IC170717A & 57951.8177 & $208.39^{+1.67}_{-1.19}$ & $25.16^{+1.41}_{-1.35}$ \\ 
 82 & IC170803A & 57968.0838 & $1.10^{+4.48}_{-1.76}$ & $4.63^{+0.41}_{-0.41}$ \\ 
 83 & IC170809A & 57974.5971 & $21.27^{+0.75}_{-1.06}$ & $-2.28^{+0.60}_{-0.67}$ \\ 
 84 & IC170824A & 57989.5538 & $41.92^{+3.04}_{-3.56}$ & $12.37^{+1.45}_{-1.30}$ \\ 
 85 & IC170922A & 58018.8712 & $77.43^{+1.14}_{-0.75}$ & $5.79^{+0.64}_{-0.41}$ \\ 
 86 & IC170923A & 58019.0213 & $173.45^{+2.38}_{-2.55}$ & $-2.54^{+0.90}_{-1.30}$ \\ 
 87 & IC171015A & 58041.0656 & $162.91^{+2.98}_{-1.72}$ & $-15.48^{+1.62}_{-1.99}$ \\ 
 88 & IC171106A & 58063.7775 & $340.14^{+0.61}_{-0.62}$ & $7.44^{+0.31}_{-0.26}$ \\ 
 89 & IC180123A & 58141.6771 & $77.12^{+2.51}_{-2.90}$ & $8.01^{+0.41}_{-0.49}$ \\ 
 90 & IC180410A & 58218.7768 & $218.50^{+0.79}_{-1.28}$ & $0.56^{+0.75}_{-0.71}$ \\ 
 91 & IC180417A & 58225.2785 & $305.73^{+3.60}_{-1.58}$ & $-4.41^{+0.68}_{-0.74}$ \\ 
 92 & IC180908A& 58369.8330 & $144.98^{+1.49}_{-2.20}$ & $-2.39^{+1.16}_{-1.12}$ \\ 
 93 & IC181023A & 58414.6927 & $270.18^{+1.89}_{-1.72}$ & $-8.42^{+1.13}_{-1.55}$ \\ 
 94 & IC181120A & 58442.7087 & $25.71^{+5.54}_{-5.28}$ & $11.72^{+2.41}_{-4.50}$ \\ 
 95 & IC181121A & 58443.5800 & $132.19^{+7.34}_{-6.99}$ & $32.93^{+4.19}_{-3.57}$ \\ 
 96 & IC190124A & 58507.1555 & $307.44^{+0.53}_{-1.14}$ & $-32.22^{+0.96}_{-0.31}$ \\ 
 97 & IC190214A & 58528.6727 & $228.25^{+0.79}_{-0.53}$ & $-4.14^{+0.37}_{-0.30}$ \\ 
 98 & IC190221A & 58535.3512 & $268.59^{+1.41}_{-1.58}$ & $-17.00^{+1.25}_{-0.50}$ \\ 
 99 & IC190503A & 58606.7244 & $120.19^{+0.66}_{-0.66}$ & $6.43^{+0.68}_{-0.75}$ \\ 
 100 & IC190515A & 58618.4506 & $127.88^{+0.79}_{-0.83}$ & $12.60^{+0.49}_{-0.46}$ \\ 
 101 & IC190613A & 58647.8294 & $312.19^{+0.66}_{-0.79}$ & $26.57^{+0.75}_{-0.71}$ \\ 
 102 & IC190619A & 58653.5516 & $343.52^{+4.13}_{-3.16}$ & $10.28^{+2.01}_{-2.76}$ \\ 
 103 & IC190730A & 58694.8685 & $226.14^{+1.28}_{-1.97}$ & $10.77^{+1.03}_{-1.18}$ \\ 
 104 & IC190922A & 58748.4047 & $167.30^{+2.81}_{-2.72}$ & $-22.27^{+3.39}_{-3.30}$ \\ 
 105 & IC190922B & 58748.9611 & $5.71^{+1.19}_{-1.27}$ & $-1.53^{+0.90}_{-0.78}$ \\ 
 106 & IC191001A & 58757.8398 & $313.99^{+6.94}_{-2.46}$ & $12.79^{+1.65}_{-1.64}$ \\ 
 107 & IC191119A & 58806.0427 & $229.31^{+5.49}_{-4.97}$ & $3.77^{+2.47}_{-2.24}$ \\ 
 108 & IC200109A & 58857.9873 & $165.45^{+3.61}_{-4.39}$ & $11.80^{+1.18}_{-1.30}$ \\ 
 109 & IC200530A & 58999.3295 & $255.37^{+2.46}_{-2.55}$ & $26.61^{+2.32}_{-3.25}$ \\ 
 110 & IC200615A & 59015.6176 & $142.95^{+1.15}_{-1.40}$ & $3.66^{+1.16}_{-1.01}$ \\ 
 111 & IC200926A & 59118.3293 & $96.46^{+0.70}_{-0.53}$ & $-4.33^{+0.60}_{-0.75}$ \\ 
 112 & IC200929A & 59121.7421 & $29.53^{+0.53}_{-0.53}$ & $3.47^{+0.71}_{-0.34}$ \\ 
 113 & IC201007A & 59129.9179 & $265.17^{+0.48}_{-0.49}$ & $5.34^{+0.30}_{-0.19}$ \\ 
 114 & IC201114A & 59167.6288 & $105.73^{+0.93}_{-1.27}$ & $5.87^{+1.01}_{-1.05}$ \\ 
 115 & IC201115A & 59168.0885 & $195.12^{+1.23}_{-1.45}$ & $1.38^{+1.27}_{-1.08}$ \\ 
 116 & IC201130A & 59183.8485 & $30.54^{+1.10}_{-1.27}$ & $-12.10^{+1.14}_{-1.11}$ \\ 
 117 & IC201209A & 59192.4276 & $6.86^{+1.01}_{-1.19}$ & $-9.25^{+0.94}_{-1.10}$ \\ 
 118 & IC201221A & 59204.5256 & $261.69^{+2.28}_{-2.46}$ & $41.81^{+1.25}_{-1.14}$ \\ 
 119 & IC201222A & 59205.0391 & $206.37^{+0.88}_{-0.75}$ & $13.44^{+0.54}_{-0.35}$ \\ 
 120 & IC210210A & 59255.4958 & $206.06^{+1.40}_{-0.95}$ & $4.78^{+0.62}_{-0.56}$ \\ 
 121 & IC210811A & 59437.0852 & $270.79^{+1.07}_{-1.08}$ & $25.28^{+0.79}_{-0.84}$ \\ 
 122 & IC210922A & 59479.7620 & $60.73^{+0.88}_{-0.61}$ & $-4.18^{+0.37}_{-0.53}$ \\ 
 
    \tableline
    \label{tab:alert table}
\end{longtable}
\end{center}

\newpage
\section{Result tables}

\begin{center}
    \begin{longtable}{l l c c c c c c c} 
    \caption[Results individual time-integrated search]{Results of the individual time-integrated analysis sorted by significance. The first column contains the index of the alert+ event as in Table \ref{tab:alert table}. The two following columns list the best-fit position of this work. The fourth and fifth columns contain the best-fit parameter of the likelihood optimization $\hat{n}_s$ and $\hat{\gamma}$. The sixth column shows the local p-values, and the seventh column the 90\% confidence level upper flux limits. The central 90\% quantiles of neutrino energies of the detected simulated events for computing the flux limit are listed in columns eight and nine. They define the range in which the flux limit is valid. The global p-value for the time-integrated single-source search is 0.98.}\\
    \tableline
    Index & R.A. [deg] & Dec [deg] & $\hat{n}_{S}$ & $\hat{\gamma}$ & $p_{local}$ & $\Phi_{90\%100
    \rm{TeV}}$ & $E_{\nu, \Phi, \text{min}}$ [TeV] & $E_{\nu, \Phi, \text{max}}$ [TeV]  \\ 
    &&&&&& [TeV cm$^{2}$ s]$^{-1}$ & & \\
    \tableline
     \endfirsthead

        \multicolumn{9}{c}{{\tablename\ \thetable{} -- continued from previous page}} \\
        \tableline
        Index & R.A. [deg] & Dec [deg] & $\hat{n}_{S}$ & $\hat{\gamma}$ & $p_{local}$ & $\Phi_{90\%100
        \rm{TeV}}$ & $E_{\nu, \Phi, \text{min}}$ [TeV] & $E_{\nu, \Phi, \text{max}}$ [TeV]\\ 
        &&&&&& [TeV cm$^{2}$ s]$^{-1}$ & & \\ 
        \tableline
        \endhead

        \multicolumn{9}{c}{{Continued on next page}} \\ 
        \endfoot

        \endlastfoot

13  &  137.87  &  -2.69  &  37.50  &  3.20  &  0.02  &  $6.88 \times 10^{-17}$  &  0.9 & 483.1 \\ 
106  &  318.48  &  11.88  &  13.38  &  2.08  &  0.03  &  $9.00 \times 10^{-17}$  &  0.7 & 132.7 \\ 
14  &  237.00  &  19.41  &  46.18  &  4.00  &  0.03  &  $9.20 \times 10^{-17}$  &  0.7 & 88.3 \\ 
2  &  298.74  &  11.74  &  38.86  &  4.00  &  0.04  &  $8.95 \times 10^{-17}$  &  0.7 & 134.0 \\ 
83  &  22.02  &  -2.13  &  28.76  &  2.93  &  0.04  &  $5.00 \times 10^{-17}$  &  0.9 & 439.5 \\ 
54  &  54.99  &  33.66  &  34.34  &  4.00  &  0.04  &  $5.58 \times 10^{-17}$  &  0.7 & 50.8 \\ 
16  &  68.36  &  40.82  &  4.48  &  1.71  &  0.05  &  $7.31 \times 10^{-17}$  &  0.7 & 40.8 \\ 
50  &  50.69  &  -0.44  &  43.72  &  4.00  &  0.06  &  $5.77 \times 10^{-17}$  &  0.8 & 380.2 \\ 
74  &  79.40  &  2.75  &  20.01  &  2.49  &  0.06  &  $6.35 \times 10^{-17}$  &  0.8 & 293.8 \\ 
10  &  284.83  &  3.32  &  16.71  &  2.38  &  0.07  &  $7.66 \times 10^{-17}$  &  0.8 & 286.4 \\ 
60  &  75.38  &  12.87  &  13.39  &  2.17  &  0.07  &  $4.77 \times 10^{-17}$  &  0.7 & 118.6 \\ 
23  &  237.76  &  19.08  &  37.66  &  4.00  &  0.07  &  $6.31 \times 10^{-17}$  &  0.7 & 95.1 \\ 
44  &  349.58  &  -13.17  &  12.27  &  2.64  &  0.07  &  $6.17 \times 10^{-17}$  &  17.3 & 5093.3 \\ 
25  &  171.74  &  26.44  &  11.18  &  2.31  &  0.07  &  $6.01 \times 10^{-17}$  &  0.7 & 67.3 \\ 
72  &  26.38  &  9.55  &  9.84  &  2.13  &  0.08  &  $2.15 \times 10^{-16}$  &  0.8 & 158.1 \\ 
45  &  350.01  &  19.02  &  40.34  &  3.91  &  0.10  &  $7.68 \times 10^{-17}$  &  0.7 & 101.2 \\ 
40  &  190.68  &  -2.35  &  26.39  &  4.00  &  0.11  &  $4.11 \times 10^{-17}$  &  0.9 & 462.4 \\ 
90  &  218.32  &  -0.15  &  32.53  &  3.49  &  0.11  &  $4.22 \times 10^{-17}$  &  0.8 & 368.1 \\ 
11  &  307.86  &  1.36  &  11.50  &  2.42  &  0.14  &  $3.24 \times 10^{-17}$  &  0.8 & 353.2 \\ 
93  &  269.42  &  -7.48  &  15.52  &  2.95  &  0.15  &  $8.71 \times 10^{-17}$  &  2.2 & 2118.4 \\ 
65  &  151.55  &  15.98  &  30.25  &  3.05  &  0.16  &  $5.88 \times 10^{-17}$  &  0.7 & 105.0 \\ 
77  &  99.20  &  -15.86  &  11.40  &  3.81  &  0.16  &  $2.11 \times 10^{-16}$  &  22.7 & 6823.4 \\ 
84  &  43.34  &  12.18  &  40.03  &  3.49  &  0.16  &  $7.54 \times 10^{-17}$  &  0.7 & 131.2 \\ 
85  &  77.43  &  5.38  &  16.73  &  2.54  &  0.17  &  $4.18 \times 10^{-17}$  &  0.8 & 237.1 \\ 
99  &  120.35  &  6.05  &  24.65  &  2.81  &  0.20  &  $4.68 \times 10^{-17}$  &  0.8 & 223.4 \\ 
55  &  133.77  &  27.71  &  22.04  &  4.00  &  0.20  &  $3.85 \times 10^{-17}$  &  0.7 & 64.4 \\ 
120  &  206.26  &  4.41  &  26.47  &  2.89  &  0.20  &  $3.79 \times 10^{-17}$  &  0.8 & 265.5 \\ 
4  &  141.25  &  47.32  &  17.24  &  2.61  &  0.21  &  $7.67 \times 10^{-17}$  &  0.7 & 32.4 \\ 
9  &  111.36  &  -0.37  &  25.89  &  4.00  &  0.21  &  $2.89 \times 10^{-17}$  &  0.8 & 371.5 \\ 
81  &  208.19  &  25.69  &  5.14  &  1.86  &  0.22  &  $5.56 \times 10^{-17}$  &  0.7 & 65.9 \\ 
28  &  70.62  &  19.43  &  8.08  &  2.09  &  0.22  &  $4.92 \times 10^{-17}$  &  0.7 & 86.5 \\ 
15  &  273.27  &  36.20  &  22.40  &  2.72  &  0.22  &  $6.14 \times 10^{-17}$  &  0.7 & 43.9 \\ 
30  &  171.12  &  27.73  &  32.96  &  3.94  &  0.23  &  $5.37 \times 10^{-17}$  &  0.7 & 62.2 \\ 
31  &  353.91  &  -1.20  &  20.58  &  2.65  &  0.24  &  $3.11 \times 10^{-17}$  &  0.9 & 394.5 \\ 
18  &  9.38  &  7.59  &  5.54  &  2.09  &  0.24  &  $4.38 \times 10^{-17}$  &  0.8 & 181.1 \\ 
115  &  194.76  &  2.47  &  17.33  &  2.48  &  0.26  &  $2.81 \times 10^{-17}$  &  0.8 & 300.6 \\ 
33  &  93.74  &  14.17  &  18.68  &  2.58  &  0.26  &  $3.76 \times 10^{-17}$  &  0.7 & 116.7 \\ 
113  &  265.17  &  5.15  &  21.84  &  4.00  &  0.28  &  $4.00 \times 10^{-17}$  &  0.8 & 248.3 \\ 
47  &  27.54  &  2.74  &  36.77  &  3.79  &  0.28  &  $3.52 \times 10^{-17}$  &  0.8 & 293.8 \\ 
34  &  225.59  &  -4.09  &  14.38  &  2.78  &  0.30  &  $3.24 \times 10^{-17}$  &  1.0 & 642.7 \\ 
119  &  206.90  &  13.27  &  20.54  &  3.16  &  0.30  &  $3.12 \times 10^{-17}$  &  0.7 & 118.6 \\ 
112  &  29.35  &  3.30  &  22.47  &  3.12  &  0.30  &  $3.70 \times 10^{-17}$  &  0.8 & 281.2 \\ 
114  &  104.46  &  6.38  &  28.34  &  3.01  &  0.30  &  $2.82 \times 10^{-17}$  &  0.8 & 224.4 \\ 
94  &  26.90  &  7.81  &  47.41  &  4.00  &  0.30  &  $7.00 \times 10^{-17}$  &  0.8 & 170.6 \\ 
7  &  330.73  &  11.10  &  21.11  &  2.90  &  0.31  &  $3.57 \times 10^{-17}$  &  0.7 & 134.0 \\ 
56  &  277.48  &  29.41  &  25.41  &  2.83  &  0.32  &  $2.86 \times 10^{-17}$  &  0.7 & 55.0 \\ 
75  &  46.95  &  15.99  &  32.99  &  3.15  &  0.32  &  $4.65 \times 10^{-17}$  &  0.7 & 106.7 \\ 
59  &  195.23  &  20.14  &  8.95  &  2.09  &  0.33  &  $6.66 \times 10^{-17}$  &  0.7 & 83.0 \\ 
79  &  283.21  &  9.37  &  32.19  &  3.43  &  0.33  &  $5.19 \times 10^{-17}$  &  0.8 & 167.1 \\ 
78  &  227.78  &  30.25  &  28.65  &  4.00  &  0.33  &  $6.23 \times 10^{-17}$  &  0.7 & 54.3 \\ 
122  &  60.27  &  -3.99  &  6.55  &  4.00  &  0.34  &  $5.41 \times 10^{-17}$  &  1.0 & 619.4 \\ 
52  &  91.86  &  12.32  &  23.31  &  3.25  &  0.34  &  $3.58 \times 10^{-17}$  &  0.7 & 130.0 \\ 
86  &  171.49  &  -2.36  &  8.32  &  2.19  &  0.34  &  $3.72 \times 10^{-17}$  &  0.9 & 467.7 \\ 
49  &  168.88  &  -1.43  &  20.41  &  4.00  &  0.35  &  $2.03 \times 10^{-17}$  &  0.9 & 409.3 \\ 
109  &  255.82  &  26.80  &  26.23  &  2.85  &  0.36  &  $2.66 \times 10^{-17}$  &  0.7 & 67.8 \\ 
88  &  340.75  &  7.44  &  19.33  &  4.00  &  0.37  &  $8.03 \times 10^{-17}$  &  0.8 & 187.5 \\ 
70  &  237.60  &  1.14  &  41.83  &  4.00  &  0.38  &  $4.43 \times 10^{-17}$  &  0.8 & 334.2 \\ 
87  &  164.10  &  -14.76  &  11.78  &  3.90  &  0.39  &  $5.17 \times 10^{-17}$  &  20.7 & 6368.0 \\ 
46  &  105.48  &  1.66  &  25.18  &  2.74  &  0.40  &  $1.73 \times 10^{-16}$  &  0.8 & 320.6 \\ 
102  &  343.52  &  9.69  &  42.18  &  3.38  &  0.40  &  $2.69 \times 10^{-17}$  &  0.7 & 158.9 \\ 
66  &  354.25  &  1.40  &  27.02  &  4.00  &  0.41  &  $2.40 \times 10^{-17}$  &  0.8 & 338.1 \\ 
92  &  144.38  &  -3.14  &  18.45  &  4.00  &  0.42  &  $2.90 \times 10^{-17}$  &  0.9 & 517.6 \\ 
100  &  128.67  &  12.76  &  5.42  &  2.18  &  0.43  &  $3.28 \times 10^{-17}$  &  0.7 & 123.0 \\ 
110  &  143.14  &  3.32  &  21.32  &  2.82  &  0.46  &  $3.36 \times 10^{-17}$  &  0.8 & 281.2 \\ 
97  &  228.84  &  -3.96  &  6.79  &  4.00  &  0.46  &  $3.30 \times 10^{-17}$  &  1.0 & 620.9 \\ 
91  &  304.94  &  -4.97  &  12.09  &  4.00  &  0.47  &  $1.34 \times 10^{-16}$  &  1.1 & 871.0 \\ 
63  &  264.13  &  -15.07  &  8.15  &  3.88  &  0.47  &  $4.46 \times 10^{-17}$  &  21.1 & 6295.1 \\ 
101  &  312.19  &  25.86  &  7.93  &  2.24  &  0.47  &  $1.57 \times 10^{-16}$  &  0.7 & 68.9 \\ 
98  &  269.80  &  -16.11  &  9.81  &  3.67  &  0.47  &  $4.83 \times 10^{-17}$  &  23.3 & 6982.3 \\ 
108  &  167.45  &  12.39  &  31.98  &  3.15  &  0.48  &  $5.15 \times 10^{-17}$  &  0.7 & 127.1 \\ 
5  &  306.96  &  19.44  &  15.41  &  2.38  &  0.49  &  $4.10 \times 10^{-17}$  &  0.7 & 86.9 \\ 
26  &  329.57  &  1.82  &  19.70  &  4.00  &  0.49  &  $1.85 \times 10^{-17}$  &  0.8 & 308.3 \\ 
68  &  122.25  &  -0.34  &  18.24  &  3.96  &  0.51  &  $2.00 \times 10^{-17}$  &  0.8 & 380.2 \\ 
37  &  301.20  &  10.50  &  5.10  &  2.17  &  0.51  &  $3.23 \times 10^{-17}$  &  0.7 & 140.9 \\ 
41  &  344.50  &  1.94  &  4.08  &  2.21  &  0.52  &  $2.54 \times 10^{-17}$  &  0.8 & 309.0 \\ 
27  &  180.72  &  3.55  &  19.65  &  3.14  &  0.55  &  $2.43 \times 10^{-17}$  &  0.8 & 291.7 \\ 
69  &  200.71  &  -31.94  &  9.69  &  3.55  &  0.57  &  $2.51 \times 10^{-17}$  &  61.9 & 11967.4 \\ 
117  &  5.67  &  -9.06  &  7.22  &  2.95  &  0.57  &  $4.89 \times 10^{-16}$  &  5.5 & 2944.4 \\ 
51  &  99.29  &  4.59  &  23.19  &  3.09  &  0.57  &  $6.52 \times 10^{-17}$  &  0.8 & 251.2 \\ 
6  &  266.87  &  13.40  &  9.33  &  2.66  &  0.57  &  $3.08 \times 10^{-17}$  &  0.7 & 118.3 \\ 
39  &  289.16  &  -14.21  &  6.66  &  3.15  &  0.59  &  $9.48 \times 10^{-17}$  &  19.1 & 5701.6 \\ 
64  &  312.60  &  59.86  &  9.68  &  2.06  &  0.59  &  $8.38 \times 10^{-17}$  &  0.6 & 23.8 \\ 
73  &  187.17  &  -6.89  &  9.57  &  2.26  &  0.61  &  $4.64 \times 10^{-17}$  &  1.6 & 1706.1 \\ 
25  &  199.80  &  32.58  &  16.22  &  2.75  &  0.61  &  $7.43 \times 10^{-17}$  &  0.7 & 51.9 \\ 
80  &  230.24  &  23.91  &  16.44  &  2.66  &  0.62  &  $4.83 \times 10^{-17}$  &  0.7 & 64.9 \\ 
105  &  4.80  &  -1.92  &  16.74  &  2.82  &  0.62  &  $2.82 \times 10^{-17}$  &  0.9 & 424.6 \\ 
61  &  261.34  &  -2.58  &  9.54  &  2.82  &  0.67  &  $2.46 \times 10^{-17}$  &  0.9 & 460.3 \\ 
104  &  166.88  &  -20.47  &  5.88  &  2.16  &  0.68  &  $2.72 \times 10^{-17}$  &  36.8 & 8851.2 \\ 
22  &  37.34  &  18.88  &  27.49  &  3.83  &  0.70  &  $2.80 \times 10^{-16}$  &  0.7 & 101.9 \\ 
32  &  169.25  &  20.84  &  22.64  &  2.87  &  0.71  &  $4.67 \times 10^{-17}$  &  0.7 & 79.4 \\ 
107  &  233.04  &  3.02  &  35.13  &  3.54  &  0.73  &  $4.45 \times 10^{-17}$  &  0.8 & 281.8 \\ 
118  &  263.47  &  42.52  &  6.25  &  1.99  &  0.74  &  $4.06 \times 10^{-17}$  &  0.7 & 40.2 \\ 
58  &  193.67  &  -3.81  &  6.84  &  4.00  &  0.74  &  $3.21 \times 10^{-17}$  &  1.0 & 588.8 \\ 
20  &  266.48  &  -5.10  &  9.66  &  3.55  &  0.74  &  $6.92 \times 10^{-17}$  &  1.1 & 948.4 \\ 
116  &  30.72  &  -11.91  &  6.42  &  3.38  &  0.76  &  $3.81 \times 10^{-17}$  &  13.6 & 4786.3 \\ 
12  &  115.64  &  -10.72  &  1.53  &  2.35  &  0.76  &  $4.62 \times 10^{-17}$  &  10.7 & 4064.4 \\ 
95  &  127.53  &  35.53  &  26.45  &  2.65  &  0.77  &  $9.04 \times 10^{-17}$  &  0.7 & 47.3 \\ 
3  &  347.90  &  22.20  &  33.17  &  4.00  &  0.79  &  $8.66 \times 10^{-17}$  &  0.7 & 78.2 \\ 
121  &  270.36  &  25.11  &  13.05  &  4.00  &  0.79  &  $4.86 \times 10^{-17}$  &  0.7 & 68.2 \\ 
38  &  284.97  &  19.11  &  19.88  &  4.00  &  0.80  &  $3.57 \times 10^{-16}$  &  0.7 & 94.0 \\ 
52  &  326.72  &  27.30  &  7.78  &  2.44  &  0.80  &  $4.65 \times 10^{-17}$  &  0.7 & 62.8 \\ 
89  &  79.63  &  8.01  &  19.75  &  4.00  &  0.81  &  $4.12 \times 10^{-17}$  &  0.8 & 166.7 \\ 
96  &  307.97  &  -32.03  &  2.48  &  2.49  &  0.81  &  $3.93 \times 10^{-17}$  &  61.5 & 12416.5 \\ 
47  &  110.83  &  11.45  &  4.77  &  3.57  &  0.82  &  $6.09 \times 10^{-17}$  &  0.7 & 136.8 \\ 
57  &  103.41  &  3.96  &  2.01  &  2.14  &  0.82  &  $3.55 \times 10^{-17}$  &  0.8 & 281.8 \\ 
62  &  79.41  &  5.00  &  12.34  &  3.04  &  0.83  &  $2.81 \times 10^{-17}$  &  0.8 & 260.0 \\ 
42  &  293.71  &  33.32  &  7.76  &  2.89  &  0.83  &  $2.83 \times 10^{-17}$  &  0.7 & 48.3 \\ 
82  &  4.61  &  4.36  &  16.46  &  4.00  &  0.85  &  $3.26 \times 10^{-17}$  &  0.8 & 272.3 \\ 
19  &  194.36  &  9.59  &  25.50  &  4.00  &  0.86  &  $4.15 \times 10^{-17}$  &  0.7 & 158.1 \\ 
103  &  225.75  &  10.77  &  12.59  &  2.88  &  0.88  &  $2.55 \times 10^{-17}$  &  0.7 & 141.6 \\ 
36  &  33.79  &  10.09  &  2.16  &  2.01  &  0.91  &  $4.01 \times 10^{-17}$  &  0.8 & 158.9 \\ 
76  &  310.75  &  9.07  &  9.19  &  2.17  &  0.93  &  $3.58 \times 10^{-17}$  &  0.7 & 165.6 \\ 
21  &  222.47  &  0.89  &  17.33  &  3.34  &  0.94  &  $2.55 \times 10^{-17}$  &  0.8 & 342.8 \\ 
67  &  214.31  &  -0.89  &  8.92  &  3.32  &  0.94  &  $5.71 \times 10^{-17}$  &  0.8 & 363.1 \\ 
71  &  193.07  &  37.50  &  14.94  &  3.19  &  0.94  &  $3.45 \times 10^{-17}$  &  0.7 & 42.6 \\ 
17  &  314.47  &  8.39  &  30.90  &  3.32  &  0.99  &  $7.36 \times 10^{-17}$  &  0.8 & 171.0 \\ 
111  &  96.46  &  -5.08  &  0.55  &  3.35  &  1.00  &  $2.74 \times 10^{-17}$  &  1.1 & 924.7 \\ 
53  &  327.74  &  5.82  &  1.69  &  3.75  &  1.00  &  $4.50 \times 10^{-17}$  &  0.8 & 229.1 \\ 
29  &  204.43  &  -2.47  &  0.00  &  2.56  &  1.00  &  $1.87 \times 10^{-17}$  &  0.9 & 465.6 \\ 
1  &  29.32  &  1.12  &  0.00  &  2.83  &  1.00  &  $1.83 \times 10^{-17}$  &  0.8 & 333.4 \\ 
8  &  88.31  &  0.33  &  0.00  &  3.20  &  1.00  &  $2.04 \times 10^{-17}$  &  0.8 & 376.7 \\ 
35  &  130.01  &  -10.69  &  0.00  &  1.50  &  1.00  &  $5.02 \times 10^{-17}$  &  10.6 & 4083.2 \\

    \tableline
    \label{tab:results_timeint}
\end{longtable}
\end{center}

\begin{center}
    \begin{longtable}{l l c c c c c c} 
    \caption[Results individual time-dependent search]{Results of the time-dependent analysis sorted by significance. The first column contains the alert+ index as in Table \ref{tab:alert table}. The next two columns list the best-fit position. The fourth and fifth columns contain the best-fit parameter of the likelihood optimization $n_s$ and $\gamma$. The sixth and seventh column list the best-fit results for the Gaussian time window with mean $\mu_T$ and width $\sigma_t$. The last column shows the local p-values. The global p-value for the time-dependent analysis is 0.156.} \\
    \tableline
    Index & R.A. [deg]& Dec [deg]& $\hat{n}_{S}$ & $\hat{\gamma}$ & $\hat{\mu}_T$ & $\hat{\sigma}_T$ & $p_{local}$ \\
    \tableline
     \endfirsthead

        \multicolumn{8}{c}{{\tablename\ \thetable{} -- continued from previous page}} \\
        \tableline
        Index & R.A. [deg]& Dec [deg]& $\hat{n}_{S}$ & $\hat{\gamma}$ & $\hat{\mu}_T$ & $\hat{\sigma}_T$ & $p_{local}$  \\
        \tableline
        \endhead

        \multicolumn{8}{c}{{Continued on next page}} \\ 
        \endfoot

        \endlastfoot

       85 & 77.43 & 5.38 & 11.98 & 2.31 & 57001 & 64 & $1.4 \times 10^{-3}$ \\ 
107 & 227.72 & 5.10 & 9.48 & 2.38 & 57774 & 9 & $1.6 \times 10^{-2}$ \\ 
17 & 318.42 & 1.75 & 10.41 & 2.45 & 57008 & 20 & $1.7 \times 10^{-2}$ \\ 
60 & 75.77 & 13.20 & 5.19 & 2.07 & 58155 & 10 & $1.9 \times 10^{-2}$ \\ 
33 & 93.74 & 14.35 & 14.45 & 2.96 & 57078 & 39 & $2.1 \times 10^{-2}$ \\ 
30 & 169.60 & 28.76 & 8.52 & 3.04 & 56153 & 5 & $2.5 \times 10^{-2}$ \\ 
83 & 21.27 & $-2.95$ & 18.44 & 2.90 & 57186 & 236 & $3.4 \times 10^{-2}$ \\ 
120 & 206.26 & 4.41 & 19.76 & 3.02 & 57427 & 103 & $5.1 \times 10^{-2}$ \\ 
99 & 120.19 & 5.87 & 12.71 & 2.68 & 56267 & 51 & $5.5 \times 10^{-2}$\\ 
87 & 164.10 & $-17.07$ & 8.10 & 3.80 & 58493 & 94 & $7.3 \times 10^{-2}$ \\ 
47 & 24.89 & 1.56 & 11.01 & 2.70 & 57764 & 22 & $7.8 \times 10^{-2}$ \\ 
23 & 238.14 & 18.42 & 8.52 & 3.03 & 57173 & 5 & $8.3 \times 10^{-2}$ \\ 
90 & 217.59 & 0.03 & 9.40 & 2.55 & 57646 & 29 & $8.4 \times 10^{-2}$ \\ 
104 & 165.83 & $-23.82$ & 6.90 & 3.12 & 58904 & 20 & $8.6 \times 10^{-2}$ \\ 
11 & 307.86 & 1.36 & 3.16 & 2.04 & 57056 & 7 & $9.5 \times 10^{-2}$ \\ 
27 & 180.53 & 3.88 & 13.77 & 3.69 & 56470 & 26 & $9.6 \times 10^{-2}$ \\ 
101 & 312.19 & 26.04 & 4.81 & 2.65 & 58692 & 5 & 0.10 \\ 
100 & 127.71 & 12.14 & 7.16 & 2.61 & 57214 & 5 & 0.11 \\ 
44 & 350.80 & $-14.90$ & 5.34 & 3.08 & 57862 & 13 & 0.11 \\ 
12 & 115.82 & $-10.53$ & 4.32 & 2.61 & 58701 & 19 & 0.11 \\ 
52 & 326.72 & 27.49 & 5.81 & 1.91 & 57677 & 5 & 0.11 \\ 
115 & 194.94 & 1.74 & 5.59 & 2.61 & 58529 & 6 & 0.12 \\ 
6 & 266.87 & 13.40 & 6.67 & 2.39 & 55551 & 37 & 0.12 \\ 
36 & 33.62 & 9.90 & 4.87 & 1.98 & 55815 & 10 & 0.12 \\ 
25 & 171.74 & 26.44 & 11.23 & 2.67 & 58063 & 120 & 0.12 \\ 
73 & 188.33 & $-6.10$ & 9.37 & 2.68 & 56698 & 78 & 0.12 \\ 
16 & 68.36 & 40.82 & 3.93 & 1.66 & 58434 & 23 & 0.13 \\ 
4 & 141.48 & 47.48 & 20.87 & 2.54 & 57931 & 268 & 0.13 \\ 
50 & 48.32 & 0.49 & 17.16 & 3.53 & 55870 & 52 & 0.19 \\ 
52 & 91.68 & 12.14 & 10.51 & 3.43 & 57694 & 12 & 0.20 \\ 
2 & 298.56 & 11.55 & 8.18 & 2.67 & 56998 & 13 & 0.22 \\ 
108 & 168.46 & 11.80 & 7.88 & 2.61 & 57075 & 7 & 0.23 \\ 
103 & 226.14 & 10.77 & 5.93 & 2.55 & 55500 & 9 & 0.25 \\ 
109 & 255.82 & 27.00 & 13.24 & 2.92 & 58940 & 32 & 0.25 \\ 
28 & 70.81 & 19.08 & 8.34 & 2.85 & 58293 & 12 & 0.25 \\ 
7 & 331.46 & 10.71 & 11.97 & 3.13 & 57753 & 18 & 0.26 \\ 
105 & 4.80 & $-0.81$ & 10.26 & 2.77 & 57854 & 21 & 0.28 \\ 
74 & 77.50 & 2.55 & 12.44 & 2.49 & 57407 & 144 & 0.29 \\ 
34 & 224.10 & $-4.09$ & 5.26 & 4.00 & 56352 & 17 & 0.32 \\ 
5 & 306.55 & 19.63 & 10.43 & 2.84 & 57728 & 37 & 0.33 \\ 
121 & 271.00 & 25.11 & 18.65 & 4.00 & 56858 & 125 & 0.33 \\ 
54 & 54.06 & 34.00 & 5.54 & 2.38 & 55465 & 8 & 0.34 \\ 
78 & 227.37 & 30.25 & 7.27 & 4.00 & 56554 & 6 & 0.34 \\ 
110 & 142.35 & 2.82 & 9.97 & 3.29 & 57011 & 10 & 0.34 \\ 
113 & 265.01 & 5.34 & 11.57 & 2.66 & 58747 & 137 & 0.35 \\ 
62 & 79.41 & 5.00 & 8.33 & 2.77 & 55881 & 23 & 0.36 \\ 
51 & 100.95 & 4.98 & 13.56 & 2.74 & 57683 & 43 & 0.36 \\ 
59 & 195.23 & 19.76 & 5.95 & 1.95 & 57777 & 43 & 0.37 \\ 
72 & 26.38 & 9.71 & 2.64 & 1.83 & 55869 & 20 & 0.38 \\ 
3 & 347.47 & 24.93 & 19.50 & 4.00 & 58817 & 67 & 0.39 \\ 
40 & 193.20 & $-3.22$ & 4.72 & 3.27 & 56069 & 5 & 0.39 \\ 
61 & 261.34 & $-2.58$ & 2.18 & 2.23 & 56311 & 5 & 0.40 \\ 
47 & 110.83 & 11.64 & 3.10 & 2.01 & 56284 & 5 & 0.40 \\ 
86 & 173.25 & $-2.54$ & 5.89 & 3.94 & 58914 & 8 & 0.40 \\ 
31 & 353.16 & $-1.40$ & 4.50 & 2.61 & 57485 & 5 & 0.41 \\ 
106 & 317.26 & 12.24 & 10.38 & 2.17 & 55648 & 194 & 0.42 \\ 
69 & 200.71 & $-31.94$ & 9.17 & 2.68 & 57641 & 149 & 0.46 \\ 
81 & 207.80 & 26.04 & 6.19 & 2.04 & 57909 & 35 & 0.47 \\ 
8 & 88.50 & 0.46 & 3.75 & 4.00 & 58729 & 5 & 0.47 \\ 
35 & 130.17 & $-10.28$ & 1.97 & 3.46 & 55727 & 5 & 0.48 \\ 
37 & 301.37 & 10.50 & 4.16 & 2.06 & 56186 & 59 & 0.48 \\ 
32 & 166.79 & 21.76 & 17.27 & 3.06 & 58039 & 105 & 0.48 \\ 
26 & 329.40 & 1.12 & 10.31 & 4.00 & 57346 & 20 & 0.49 \\ 
96 & 307.97 & $-32.03$ & 5.75 & 2.71 & 58223 & 130 & 0.49 \\ 
55 & 133.55 & 27.71 & 12.73 & 4.00 & 57202 & 96 & 0.49 \\ 
65 & 151.05 & 14.93 & 10.71 & 2.69 & 56073 & 74 & 0.50 \\ 
102 & 341.35 & 11.01 & 23.55 & 3.50 & 55932 & 225 & 0.50 \\ 
53 & 327.92 & 5.82 & 7.41 & 3.54 & 57307 & 5 & 0.51 \\ 
10 & 284.83 & 3.32 & 2.92 & 1.95 & 58860 & 51 & 0.52 \\ 
1 & 29.51 & 1.23 & 4.52 & 4.00 & 58152 & 5 & 0.52 \\ 
97 & 228.25 & $-4.44$ & 4.67 & 4.00 & 57925 & 23 & 0.54 \\ 
18 & 9.38 & 7.59 & 2.45 & 1.76 & 56920 & 133 & 0.55 \\ 
114 & 105.55 & 6.38 & 10.56 & 3.26 & 57438 & 73 & 0.55 \\ 
98 & 268.20 & $-16.29$ & 8.79 & 3.93 & 55171 & 32 & 0.56 \\ 
75 & 47.55 & 15.44 & 6.99 & 2.85 & 56015 & 6 & 0.56 \\ 
19 & 199.21 & 8.87 & 9.84 & 3.33 & 58512 & 16 & 0.58 \\ 
21 & 223.65 & 1.67 & 6.06 & 2.04 & 57754 & 58 & 0.59 \\ 
46 & 105.67 & 0.97 & 8.75 & 2.52 & 56044 & 62 & 0.60 \\ 
49 & 168.88 & $-1.43$ & 12.14 & 4.00 & 57034 & 135 & 0.60 \\ 
88 & 340.60 & 7.59 & 6.69 & 4.00 & 56225 & 10 & 0.63 \\ 
122 & 60.12 & $-3.99$ & 4.24 & 4.00 & 57539 & 33 & 0.64 \\ 
20 & 266.67 & $-5.10$ & 9.44 & 3.81 & 57444 & 122 & 0.65 \\ 
79 & 282.81 & 8.26 & 16.67 & 2.98 & 57871 & 144 & 0.66 \\ 
92 & 144.78 & $-2.95$ & 12.23 & 4.00 & 58406 & 95 & 0.67 \\ 
91 & 307.53 & $-4.97$ & 6.03 & 2.39 & 57105 & 26 & 0.68 \\ 
95 & 130.33 & 36.92 & 10.91 & 4.00 & 55469 & 51 & 0.68 \\ 
82 & 359.34 & 4.36 & 11.18 & 3.54 & 56650 & 26 & 0.68 \\ 
80 & 230.45 & 23.54 & 8.44 & 2.91 & 56927 & 13 & 0.68 \\ 
112 & 29.00 & 3.47 & 3.64 & 4.00 & 57045 & 5 & 0.69 \\ 
93 & 270.56 & $-7.29$ & 6.61 & 2.10 & 58532 & 44 & 0.69 \\ 
70 & 237.20 & 1.14 & 16.00 & 3.27 & 56123 & 65 & 0.70 \\ 
22 & 37.14 & 18.69 & 16.67 & 2.89 & 57494 & 112 & 0.73 \\ 
66 & 354.64 & 1.73 & 4.05 & 2.30 & 57815 & 5 & 0.73 \\ 
119 & 207.07 & 13.27 & 6.99 & 3.26 & 56078 & 20 & 0.74 \\ 
25 & 199.80 & 32.58 & 6.94 & 2.41 & 55763 & 38 & 0.74 \\ 
39 & 289.35 & $-15.33$ & 3.13 & 3.60 & 58116 & 8 & 0.75 \\ 
63 & 263.76 & $-15.07$ & 5.68 & 3.72 & 57964 & 136 & 0.76 \\ 
15 & 272.55 & 36.20 & 5.39 & 1.97 & 55850 & 40 & 0.76 \\ 
57 & 102.85 & 3.77 & 5.77 & 4.00 & 57412 & 5 & 0.77 \\ 
77 & 97.34 & $-15.06$ & 2.86 & 3.94 & 57966 & 5 & 0.79 \\ 
111 & 96.46 & $-5.08$ & 3.54 & 3.78 & 55790 & 14 & 0.80 \\ 
117 & 6.66 & $-9.98$ & 9.12 & 3.27 & 55227 & 69 & 0.82 \\ 
94 & 25.91 & 7.61 & 24.88 & 3.40 & 56859 & 99 & 0.82 \\ 
84 & 39.94 & 13.64 & 23.44 & 4.00 & 58516 & 221 & 0.83 \\ 
14 & 235.34 & 19.76 & 10.65 & 2.42 & 58482 & 76 & 0.84 \\ 
58 & 194.20 & $-3.81$ & 6.96 & 2.76 & 55443 & 224 & 0.84 \\ 
89 & 77.12 & 7.68 & 4.52 & 4.00 & 57329 & 5 & 0.84 \\ 
76 & 310.55 & 6.39 & 9.45 & 2.80 & 55956 & 23 & 0.85 \\ 
67 & 214.12 & $-0.71$ & 6.49 & 3.58 & 57509 & 22 & 0.87 \\ 
29 & 205.30 & $-2.65$ & 6.59 & 3.19 & 56484 & 7 & 0.87 \\ 
64 & 312.23 & 60.79 & 5.30 & 2.32 & 56903 & 7 & 0.88 \\ 
13 & 137.87 & $-2.87$ & 20.28 & 3.26 & 57963 & 386 & 0.90 \\ 
9 & 111.16 & $-0.21$ & 6.61 & 2.97 & 58059 & 19 & 0.90 \\ 
68 & 122.43 & $-1.08$ & 9.92 & 3.52 & 55389 & 127 & 0.91 \\ 
56 & 280.64 & 30.35 & 12.20 & 2.57 & 57865 & 153 & 0.91 \\ 
116 & 30.54 & $-11.15$ & 3.03 & 2.49 & 57403 & 5 & 0.93 \\ 
41 & 345.19 & 1.40 & 8.92 & 3.69 & 55980 & 35 & 0.93 \\ 
42 & 293.51 & 33.32 & 4.21 & 4.00 & 56102 & 5 & 0.94 \\ 
38 & 287.16 & 19.47 & 10.07 & 4.00 & 55637 & 45 & 0.96 \\ 
45 & 346.71 & 18.24 & 12.32 & 2.61 & 55259 & 143 & 0.97 \\ 
118 & 263.21 & 42.52 & 6.51 & 3.44 & 56064 & 6 & 0.98 \\ 
71 & 191.42 & 35.21 & 4.87 & 1.60 & 55301 & 42 & 0.99 \\  

    \tableline
    \label{tab:results_timedep}
\end{longtable}
\end{center}

\bibliography{references}{}
\bibliographystyle{aasjournal}



\end{document}